\documentclass[AMA,STIX1COL,svgnames]{WileyNJD-v2}

\usepackage[framemethod=TikZ]{mdframed}
\usepackage{amsmath,fixmath,listing,color,mathastext,graphicx,wrapfig,amssymb}
\usepackage{tikz}
\usetikzlibrary{positioning}
\usetikzlibrary[shapes.misc]
\usetikzlibrary{arrows,automata}
\usepackage{todonotes}
\usepackage{threeparttable}
\usepackage{hyperref}
\hypersetup{colorlinks=true, linkcolor=blue, filecolor=blue, urlcolor=blue, citecolor=black,}
\usepackage{color}
\usepackage{listings}
\usepackage{xkeyval} 
\input{listings-stata.tex}

\articletype{Research article}
\received{XXXX-XX-XX}  
\revised{XXXX-XX-XX}
\accepted{XXXX-XX-XX}
\begin{document}
\lstset{style=stata-editor, 
basicstyle=\footnotesize\ttfamily, 
numbers=left, 
numberstyle=\tiny\color{gray}, 
stepnumber=1, 
numbersep=5pt, 
backgroundcolor=\color{white}, 
showspaces=false, 
showstringspaces=false, 
showtabs=false, 
frame=single, 
rulecolor=\color{black}, 
tabsize=1, 
captionpos=b, 
breaklines=true, 
extendedchars=true,
}
\title{Tutorial: Introduction to computational causal inference using reproducible Stata, R and Python code}

\author[a]{Matthew J. Smith}
\author[a]{Camille Maringe}
\author[a]{Bernard Rachet}
\author[b]{Mohammad A. Mansournia}
\author[c]{Paul N. Zivich}
\author[c]{Stephen R. Cole}
\author[a,d,e]{Miguel Angel Luque-Fernandez*}

\authormark{Smith and Luque-Fernandez}

\address[a]{Inequalities in Cancer Outcomes Network, Department of Non-communicable Disease Epidemiology. London School of Hygiene and Tropical Medicine, London, U.K.}
\address[b]{Department of Epidemiology and Biostatistics. Tehran University of Medical Sciences, Tehran, Iran.}
\address[c]{Department of Epidemiology. University of North Carolina at Chapel Hill, North Caroline, U.S.}
\address[d]{Non-communicable Disease and Cancer Epidemiology Group, Instituto de investigacion Biosanitaria de Granada (ibs.GRANADA), University of Granada, Granada, Spain}
\address[e]{Biomedical Network Research Centers of Epidemiology and Public Health
(CIBERESP), Madrid, Spain.}

\corres{* Miguel Angel Luque-Fernandez,
\email{miguel-angel.luque@lshtm.ac.uk}}

\presentaddress{Keppel St, Bloomsbury, London WC1E 7HT, United Kingdom}

\abstract{\textbf{Abstract}
The purpose of many health studies is to estimate the effect of an exposure on an outcome. It is not always ethical to assign an exposure to individuals in randomised controlled trials, instead observational data and appropriate study design must be used. There are major challenges with observational studies, one of which is confounding that can lead to biased estimates of the causal effects. Controlling for confounding is commonly performed by simple adjustment for measured confounders; although, often this is not enough. Recent advances in the field of causal inference have dealt with confounding by building on classical standardisation methods. However, these recent advances have progressed quickly with a relative paucity of computational-oriented applied tutorials contributing to some confusion in the use of these methods among applied researchers. In this tutorial, we show the computational implementation of different causal inference estimators from a historical  perspective where different estimators were developed to overcome the limitations of the previous one. Furthermore, we also briefly introduce the potential outcomes framework, illustrate the use of different methods using an illustration from the health care setting, and most importantly, we provide reproducible and commented code in Stata, R and Python for researchers to apply in their own observational study. The code can be accessed at \hyperlink{https://github.com/migariane/TutorialCausalInferenceEstimators}{https://github.com/migariane/TutorialCausalInferenceEstimators}
}


\maketitle

\section{Introduction}

\noindent The questions that motivate most studies in the health, social and behavioral sciences are causal in nature. For example, what is the mortality risk difference amongst patients offered a surgical procedure versus those who did not receive it in a given population?\cite{article} Causal inference methods may be used to answer this scientific question where a clinical trial is unfeasible or unethical (e.g. a surgical procedure).\cite{Rubin2007TheTrials} Causal inference methods are based on the counterfactual framework introduced by Neyman in the randomized experiment setting, then extended to time-fixed and time-varying observational studies by Rubin and Robins, respectively.\citep{Rubin1974EstimatingStudies.,Robins1986AEffect} 
\\

\noindent Over the years, rapid on-going advances in the field causal inference have created a set of different approaches to estimate the causal effect of a treatment (or exposure) on an outcome (i.e., methods that incorporate propensity scores, the G-computation, or a combination of both, namely double-robust methods). Overall, double-robust methods now are preferred over naive regression approaches because the latter are biased under misspecification of a parametric outcome model when the research question is causal in nature.\cite{vanderLaan2011TargetedData,doi:10.2105/AJPH.2018.304916}
\\

\noindent In this tutorial we introduce the most commonly used estimators in causal inference from a practical computational perspective: allowing readers to learn by using the replicable code. We use the Stata statistical software (\textit{StataCorp. 2020. Stata Statistical Software: Release 16. College Station, TX: StataCorp LLC, USA}). We also provide readers with access to code in R statistical software (\textit{R Development Core Team (2020). R: A language and environment for statistical computing. R Foundation for Statistical Computing, Vienna, Austria}) and Python statistical software (\textit{Python Software Foundation (2020)}) and a link to a GitHub repository where readers can access the data, along with all software code, to replicate our empirical example accessed at \hyperlink{https://github.com/migariane/TutorialCausalInferenceEstimators}{https://github.com/migariane/TutorialCausalInferenceEstimators}.
\\

\noindent To illustrate the computation of the most common causal inference estimators we use a classical empirical data set from the prospective cohort study of Connors \textit{et al} (1996).\citep{Connors1996ThePatients} The data is open source and available at: \href{http://biostat.mc.vanderbilt.edu/wiki/Main/DataSets}{http://biostat.mc.vanderbilt.edu/wiki/Main/DataSets}. The study evaluated the effectiveness of right heart catheterisation (RHC) on short-term mortality (30 days) of 5,735 critically ill adult patients, 2,184 treated and 3,551 controls in five United States teaching hospitals receiving care in an ICU for 1 of 9 prespecified disease categories between 1989 and 1994. In our illustration, the outcome is short-term mortality (Y) defined as 30 days after ICU admission, RHC as the main treatment or exposure (A), and in (\textit{\textbf{W}}) we include the set of confounders. 
\\

\noindent In the following sections, we will illustrate the computational implementation of different estimators for our estimand of interest: the average treatment effect (ATE), also known as the risk difference. In section two we introduce the causal language, followed by section three that details the computation of the G-formula non-parametrically and parametrically using the G-computation method. Then, in section four we introduce the methods based on the inverse probability-of-treatment weights (IPTW). Afterwards, in section five, we shortly introduce the computation of double-robust methods, and finally in section six we present the targeted maximum likelihood estimation (TMLE) framework. We illustrate the computation of the different estimators using Stata in a set of boxes containing commented code. 
\\

\section{Causal language: potential outcomes}

Commonly, the aim of an observational study is to answer a scientific question that characterises the effect of an exposure or treatment on an outcome. This question is translated to an \textit{estimand} whose estimate will provide a relevant answer to the characterisation of the effect of the exposure on the outcome (i.e., by how many times is the risk of the outcome increased, given a particular level of the exposure). On the other hand, \textit{estimators} are mathematical functions using the observed values of the observations in the sample (that is, a function of the random variables) that generates the quantitative value for the estimand. The estimators are represented by algebraic equations that explicitly describe a function of the realized observations.\\

\noindent The most common estimand in causal inference is the ATE and is the focus of this tutorial. The ATE is a function of the underlying distribution of the counterfactual outcomes, which can be estimated non-parametrically or parametrically.\cite{Robins1986AEffect} The ATE is defined by an average of the difference of two random variables (i.e., the potential outcomes $Y(1)$ and $Y(0)$).\cite{Rubin2007TheTrials, Gutman2015EstimationStudies} Thus, to describe the ATE we need to introduce the language of the \text{Potential Outcomes Framework} also known as (a.k.a) Neyman-Rubin Potential Outcomes framework.\cite{Rubin2007TheTrials} 
\\

\noindent To illustrate the framework we use the study of Connors based on intensive care medicine \textit{et al} (1996).\citep{Connors1996ThePatients} Let Y, the outcome, denote the vital status of the patient in an intensive care unit (ICU) at 30 days after admission. Let A, the exposure, denote the treatment variable of whether or not the patient received right heart catheterisation (RHC) during his stay at the ICU. Let C, a confounder, denote the gender of the patient. 
\\

\noindent In this RHC study, each patient has two \textit{potentially observed} outcomes (i.e., Y(a)), where the first is $Y(1)$ if they received RHC, and the second is $Y(0)$ if they did not receive RHC.\cite{Rubin1974EstimatingStudies.} We say \textit{potentially observed} because only one of these two outcomes can ever be observed as each patient only receives one of the treatments. As an example from table \ref{table:1}, Ronald has two potential outcomes: firstly, $Y(0) = 1$ says that if Ronald were not to receive RHC then he would die within 30 days, and secondly, $Y(1) = 0$ says that if Ronald were to receive RHC then he would not die within 30 days. The two potential outcomes mentioned thus far can also be stated as $Y(0)$ and $Y(1)$, respectively.
\\

% \begin{wraptable}{l}{7cm}
\begin{table}[ht!]
\centering
 \begin{tabular}{ c c c c c c } 
 \hline
 \textbf{Patient} & \textbf{Y} & \textbf{A} & \textbf{C} & $Y(0)$ & $Y(1)$ \\ [0.5ex] 
 \hline   
 Ronald   & 1 & 0 & 0 & 1 & 0 \\ 
 Malala   & 1 & 1 & 1 & 1 & 1 \\
 Florence & 1 & 1 & 1 & 0 & 1 \\
 Karl     & 0 & 1 & 0 & 0 & 0 \\
 Marie    & 1 & 0 & 1 & 1 & 1 \\
 Greta    & 0 & 1 & 1 & 0 & 0 \\
 Miguel   & 1 & 0 & 0 & 1 & 0 \\
 Alice    & 0 & 1 & 1 & 0 & 0 \\
 Matthew  & 1 & 1 & 0 & 1 & 1 \\
 Ada      & 0 & 0 & 1 & 0 & 0 \\ [1ex] 
 \hline
 \end{tabular}
 \caption{Potential outcomes framework: C = Binary confounder, A = Binary exposure, Y = Binary outcome}
\label{table:1}
%\end{wraptable}
\end{table}

\noindent Causal effects of a treatment on an outcome are defined in terms of a contrast between the \textit{potential outcomes} under different treatment levels (i.e. $ Y(0) $ and $ Y(1) $), as formalised by Rubin.\cite{Rubin2011CausalOutcomes} An individual causal effect exists for patient \textit{i} if $ Y_{i}(0) \neq Y_{i}(1) $. For example, there is an individual causal effect for Ronald as $ Y_{i}(0) \neq Y(1) $, but not for Karl as $ Y(0) = Y(1) = 0 $. However, only one potential outcome can ever be observed, corresponding to the treatment that was actually taken. Therefore, individual causal effects are not possible to estimate. To elaborate, in this example, the causal effect of RHC on risk of death is a comparison of the potential outcomes for patient \textit{i}, such that the ATE is defined as

    $$\text{ATE}\,=\,E[P(Y(1)\,|\,A=1) + \underbrace{P(Y(1)\,|\,A=0)]}_{Unobserved} \,-\, E[P(Y(0)\,|\,A=0) + \underbrace{P(Y(0)\,|\,A=1)]}_{Unobserved},$$ \\
    
\noindent where \textbf{E} refers to expectation (i.e., average) and \textbf{P} refers to probability. The potential outcome, $Y(1)\,|\,A=0$, is counterfactual to $Y(1)\,|\,A = 1$ because $Y(1)\,|\,A=0$ states what the outcome for patient \textit{i} (who did not receive the treatment, i.e., A = 0) would have been if the patient had counterfactually received it, i.e., A = 1. Ronald's observed outcome was death before 30 days ($Y(0)=1$) given he did not receive RHC, and his (unobserved) counterfactual outcome was alive at 30 days ($Y(1)=0$) given he did receive RHC. 
\\

\noindent Since the patient receives only $A=1$ or $A=0$, only one outcome for each patient can be observed, the counterfactual is not observed. In experimental designs, the randomisation makes the ATE estimate, i.e., E(Y|A=1) - E(Y|A=0), unbiased for the ATE estimand: 
\\
    $$\text{ATE}\,=\,E[Y(1)\,-\,Y(0)].$$ 
\\
\noindent Since randomisation makes confounding random and the expected value of the distribution of confounders averaged over allocations are the same:\citep{Greenland2015LimitationsUnfaithfulness}
\\
    $$E[P(Y(1)|A=1)] = E[P(Y(1)|A=0)]\;\text{and}\; E[P(Y(0)|A=0)] = E[P(Y(0)|A=1)].$$ 
\\
\noindent However, randomisation is often unethical in clinical trials or unfeasible in observational studies. We must make certain (untestable) assumptions to identify potential outcomes from the observed data and estimate the causal estimand.\cite{Robins1986AEffect} Given that the potential outcomes are not necessarily directly observed from the data, to identify the ATE from observable data (i.e., from Table 1) the following three assumptions are made:
\\

\noindent \textbf{1. Counterfactual consistency}

\noindent Counterfactual consistency holds if the observed outcome for all exposed individuals equals their outcome if they had been exposed, and likewise for unexposed individuals. This means that the definition of the exposure is consistent for all individuals. For example, in table 1, Ronald's observed outcome equals his potential outcome if he had not been exposed ($Y(0) = 1$). This means that the definition of the exposure is consistent for Ronald (the same applies for all the other patients). Analytically, consistency is represented by:
$$Y = A Y(1) + (1-A)Y(0)$$ 

\noindent \textbf{2. Conditional exchangeability} 

\noindent In randomised studies, conditional exchangeability, and in fact, marginal exchangeability and full exchangeability, hold because the exposed individuals, had they not been exposed, would have had the same average outcome as the unexposed, and vice versa. This cannot be guaranteed in observational studies but it can be assumed to hold if the unmeasured risk factors of the outcome are equally distributed between the exposed and the unexposed groups conditional on the measured confounders. Thus, using the language of the potential outcomes the conditional exchangeability assumption (a.k.a conditional independence, unconfoundedness or ignorability) is given: 
\\
$$Y(a) \amalg A\mid \textbf{C} \; \forall \; a \in \{0, 1\}$$ 
Hence, the conditional mean independence is given\\
$$
E[Y(a)\mid A=1, C=c]=E[Y(a)\mid A=0, C=c]=E[Y(a) \mid C=c]\; \forall a\in{0,1}
$$

\noindent \textbf{3. Positivity} 

\noindent Positivity holds if the conditional probability of being exposed (and similarly for being unexposed) is greater than zero.
Therefore, if $ P(C=c) > 0 \text{ then } P(A=a\mid \textbf{C}=c) > 0 \; \forall \; C \in \textbf{c}, a \in \{0, 1\}$ .We must also assume noninterference, meaning that the potential outcome of one individual was not influenced by the exposure of any other individual (i.e., independence of observations). 
\\

\noindent Transitioning from unobserved potential outcomes to a setting where we estimate the causal effect can be done by applying these assumptions. Suppose we are interested in estimating the ATE, and for simplicity, under the assumption there is only one confounder, namely \textit{C}.
\\
   $$\text{ATE = } E[Y(1)] \, - \, E[Y(0)]$$
\\
\noindent Note that we do have the possibility to estimate the average treatment among the treated (ATT) represented by: 
\\
$$\text{ATT}\,=\,E[Y(1) \mid A=1] - E[Y(0) \mid A=1]$$
\\
  
\noindent By the law of total probability
\\
$$P[Y(a)=1] \, = \, \sum_{c} P[Y(a)=1 \,|\, C=c]\,P(C=c)$$ 
\\
    
\noindent By conditional exchangeability the right hand side is \\

    $$\sum_{c} P[Y(a)=1 \,|\,A=a,C=c]\,P(C=c)$$ \\

\noindent This is possible since we are assuming that within levels of \textit{C} the predictors of the outcome are equally distributed between the treatment (RHC) and non-treatment (non-RHC) groups: that is we have achieved what would happen if patients were randomised to each treatment group. If we assume consistency the right hand side is \\
    $$\sum_{c} P[Y=1 \,|\,A=a,C=c]\,P(C=c)$$ \\
Thus, under these assumptions, the ATE is defined as

\begin{equation*} 
P(Y(1)=1) - P(Y(0)=1)
\end{equation*}
\begin{eqnarray}
\sum_{c} P[Y=1 \,|\,A=1,C=c]\,Pr(C=c) \,-\, \sum_{c} P[Y=1 \,|\,A=0,C=c]\,P(C=c),
\end{eqnarray}
\\
\noindent and the average treatment effect among the treated (ATT) is defined as
\\
\begin{eqnarray}
\sum_{c} P[Y=1 \,|\,A=1,C=c]\,Pr(C=c \mid A=1) \,-\, \sum_{c} P[Y=1 \,|\,A=0,C=c]\,P(C=c \mid A=1) 
\end{eqnarray}

\begin{figure}[ht!]
\centering
\scalebox{1.3}{
\begin{tikzpicture}
\draw[very thick] (-2,0) circle (10pt); %C
\draw[very thick] (0,0) circle (10pt); %A
\draw[very thick] (2,0) circle (10pt); %Y
\draw (-2,0) node[font=\footnotesize] (w) {\textbf{W}};
\draw (0,0) node (a) {A};
\draw (2,0) node (y) {Y};
%----W----%
\draw[line width=1,->] (-1.6,0) -- (-0.4,0);
\draw[->, line width=1] (-1.7,0.3) to [out=45,in=135, looseness=1.0] node[above,font=\footnotesize]{} (1.7,0.3);
%----A----%
\draw[red,line width=1,->] (0.4,0) -- (1.6,0);
\end{tikzpicture}
}
\caption{Y: mortality within 30 days ; A: right heart catheterisation; \textbf{W}: full set of confounders outlined in Connors \textit{et al.} (1996)}
\label{Figure:1}
\end{figure}
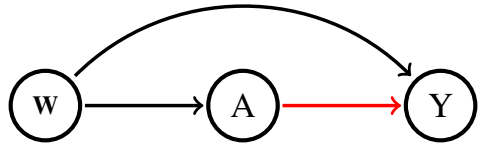

\noindent We have transitioned from the unobserved potential outcomes to a setting where we can estimate our causal estimand using the distribution of the observed data using formula (1), namely the G-formula.\citep{Robins1999AssociationModels} To do it we are going to compute different estimators of the ATE using the prospective cohort study of Connors \textit{et al} (1996).\citep{Connors1996ThePatients} Figure (\ref{Figure:1}) represents the association of the vector of predefined confounders (i.e., \textbf{W}) on the treatment (A) and the outcome (Y). 

\noindent We assume that: all the variables included in \textbf{W} are confounders of the effect of A on Y; there are no intermediate variables; and there is no residual confounding. Therefore, the set of covariates included in \textbf{W} suffices to assume the conditional mean independence to estimate the ATE.  In our example the outcome (Y) is mortality within 30 days, the treatment (A) is the right heart catheterisation, and we use patients' sex, age, education, race and the presence of a carcinoma as confounders represented by the vector \textbf{W} in Figure 1. Note, that the original data include a highly dimensional set of confounders; however, for illustrative purposes, we include five of confounders. With Connor's study as an example, and using a set of different causal inference estimators for the ATE and the ATT, we estimate the standardised short term mortality risk difference of death for those patients who were offered RCH versus those who were not offered RHC.
\\

\section{G-formula: Formal Definition and application}

\subsection{Non-parametric G-formula}

\noindent Classical epidemiological methods use regression adjustment to explain the main effect of a risk factor on a disease or outcome. Regression adjustment, controlling for confounding, requires making the assumption that the effect is constant across levels of confounders (\textbf{W}) included in the model.\cite{doi:10.2105/AJPH.2018.304916} However, in non-randomised observational studies, the effect measure is confounded given the different distribution of individual characteristics at baseline. In the instance of differential age distributions between two groups, age-standardisation to one population's age distribution, or to an external standard, is performed. The G-formula is a generalisation of the classical standardisation procedure, and it allows obtaining an unconfounded marginal estimation of the ATE that is given for a binary treatment by: \cite{Robins1986AEffect}

\begin{equation} 
\text{ATE}\,=\,\sum_{w}\,\left[P(Y = 1\mid A=1,\textbf{W}=w)\,-\,P(Y = 1\mid A=0,\textbf{W}=w)\right]P(\textbf{W}=w).
\end{equation}

\noindent where,   
\begin{equation*}
    P(Y = 1 \mid A = a, \textbf{W} = w)\,=\,\frac{P(\textbf{W} = w, A = a, Y = 1)}{\sum_{y}\,P(\textbf{W} = w, A = a, Y = y)} 
\end{equation*}
\\

\noindent is the conditional probability distribution of the outcome Y = y, given the exposure or treatment A = a, and the set of counfounders \textbf{W} = w. Note that the implementation of the G-formula only requires the use of the total law of probability and the expectation of random variables. In probability theory, the total probability is a fundamental rule relating marginal probabilities to conditional probabilities. Likewise, marginal and conditional probabilities are used to estimate the expectation of random variables.
\\

\noindent In the following set of 7 boxes we show how to compute the estimate of the effect of RHC on short term mortality using the non-parametric G-formula using Stata (the same computing approach is provided in the appendix for R and Python users). We will use gender as the unique confounder. It is an unrealistic scenario but it is made intentionally to allow readers to readily appreciate the computing implementation of the non-parametric G-formula. In contrast to the naive approach, the way we adjust for confounding based on the generalisation of standardisation (G-formula) is more consistent: we assume that the RHC effect on mortality can differ by sex while using the naive approach assumes that the effect is the same for both sexes.\citep{doi:10.2105/AJPH.2018.304916} It is a subtle difference but provides a better adjustment for confounding. We do the same for the average treatment among the treated (ATT). Sometimes, the ATT is of interest (i.e, under the presence of near positivity violations). We use the bootstrap for statistical inference and provide 95\% confidence intervals (CI) for the ATE. Finally, we also show readers how to implement the non-parametric estimation of the ATE and ATT using a regression modelling approach based on a fully saturated regression model.
\[\]

\noindent \textbf{Box 1: Setting the data} \\
\noindent Here we declare the global variables Y, A, C, and W to match the presented algebraic nomenclature. i.e., Y: outcome, A: treatment, C: one unique confounder, and W: a set of confounders. We will use these global variables throughout the implementation of the different methods.
\\
\begin{lstlisting}
    clear
    set more off
    use "rhc.dta"
    global Y death_d30                     // Outcome
    global A rhc                           // Exposure or treatment
    global C gender                        // One unique confounder of the set of W confounders
    global W gender age edu race carcinoma // A set of five confounders
\end{lstlisting}

\noindent\textbf{Box 2: Regression - a naive approach} \\
\noindent We first introduce a naive approach to estimating the ATE; we regress the outcome over the treatment and adjust for the confounder (i.e., gender). In naive regression, it is assumed that there is an equal probability of an individual being a certain level of the confounder (i.e., gender). Naively, we say there is strong evidence (p<0.001) that the risk of death within 30 days is 7.4\% higher amongst those with RHC and 95\% CI (4.84-9.86) (Table 3).
\\
\begin{lstlisting}
    regress $Y $A $C // Risk differences = 7.4%; 95% CI (4.84-9.86); p<0.001
\end{lstlisting}

\noindent\textbf{Box 3: Non-parametric G-Formula for the ATE} \\
\noindent  First, we compute the marginal probability of the confounder, saving it in a matrix "m", and generate two new variables named \textit{genderf} for females and \textit{genderm} for males (i.e., the proportion of females was 44\%, thus 56\% are males, which shows unequal proportions). Afterwards, we compute, and save in a matrix, the expected conditional probabilities of the outcome by levels of the exposure and the confounder. We substitute the results of the matrix into the G-formula, given in (2), and compute the ATE. The risk of death within 30 days is 7.4\% higher amongst those with RHC compared to those patients who did not get the RCH.\\

\begin{lstlisting}
    proportion $C           // Marginal probability of C
    matrix m=e(b)
    gen genderf = m[1,1]
    sum genderf
    gen genderm = m[1,2]
    sum genderm
    ssc install sumup
    sumup $Y, by($A $C)     // Expected conditional probabilities of the outcome by levels of A and C
        // from sumup command extract the conditional means by the given levels of A and C i.e. zero and one
    matrix y00 = r(Stat1)   // [6,1] matrix for E(Y|A=0,C=0)
    matrix y01 = r(Stat2)   // [6,1] matrix for E(Y|A=0,C=1)
    matrix y10 = r(Stat3)   // [6,1] matrix for E(Y|A=1,C=0)
    matrix y11 = r(Stat4)   // [6,1] matrix for E(Y|A=1,C=1)
        // see "matrix list y00": position subscript [3,1] is the one of interest
    // Applying the G-formula
    gen EY1 = ((y11[3,1]-y01[3,1]))*genderm
    gen EY0 = ((y10[3,1]-y00[3,1]))*genderf
    qui: mean EY1 EY0
    matrix ATE =  r(table)
    display "The ATE is: "  ATE[1,1] + ATE[1,2] // Applying the G-formula
    drop EY1 EY0
    // Also one can try
    gen ATE = ((y11[3,1]-y01[3,1]))*genderm + ((y10[3,1]-y00[3,1]))*genderf
    qui sum ATE
    drop ATE
\end{lstlisting}

\noindent Note that the results from the naive and G-formula approaches are the same (Table 3). However, the interpretation varies because the naive approach is a conditional estimation interpreted as the individual risk among those who actually had the exposure holding the levels of the confounder constant at their mean value (i.e., gender) at each combination of the levels of the exposure and outcome (i.e., instead of using the marginal probabilities of males and females as in row 15 from box 3, the mean value of gender is used). Note that it is difficult to conceptualize the mean value of a categorical variable such as gender. On the other hand, the G-formula is the contrast of the conditional mean outcome given treatment A for both male and females standardised using the marginal probability of gender (i.e., males and females) and averaged across the levels of males and females in the sample (i.e., row 15 in box 3) and therefore must be interpreted at a population level. The interpretation of the conditional treatment effect, adjusted for gender, from the naive approach becomes difficult to interpret after standardisation using the mean value of gender and requires the assumption that the effect of the treatment of interest is constant across the levels of the confounders included in the model (i.e., is the same for males and females).\citep{doi:10.2105/AJPH.2018.304916, Keil2014TheExample} However, in observational studies the effect of risk factors or treatments might differ across individuals with different susceptibilities or characteristics. Therefore, the G-Formula has became a powerful alternative to the multivariable regression adjustment to better control for confounding and evaluate the effects of treatment and exposures in observational studies. \citep{Robins1986AEffect} 
\\

\noindent So far we have calculated the ATE amongst those who had RHC or not, but often there are identifiability issues where the probability of the exposure or treatment is nearly zero or zero for some groups. Therefore, the average treatment effect amongst the treated (ATT) might be of interest as it is the marginal contrast of the potential outcomes amongst only those who had RHC. Therefore, this comes with a slightly weaker assumption of conditional mean exchangeability such that the probability of the potential outcome among those who died if, possibly contrary to fact, they had not received the treatment (i.e., RHC in our example) is independent of the exposure given the confounder.
\\

\noindent\textbf{Box 4: Non-parametric G-Formula for the ATT}\\
\noindent In box 5 we show how to compute the ATT non-parametrically based on the implementation of the G-formula. The estimate of the ATT is 7.3\% and is interpreted as the increased risk of death within 30 days amongst patients with RHC, had they not been treated.
\\

\begin{lstlisting}[escapeinside={(*}{*)}]
    // Expected conditional probability of the outcome amongst those with RHC
    proportion $C if $A==1      
    matrix m=e(b)
    gen genderfatet = m[1,1]
    gen gendermatet = m[1,2]
    // G-formula
    gen EY1 = ((y11[3,1]-y01[3,1]))*gendermatet
    gen EY0 = ((y10[3,1]-y00[3,1]))*genderfatet
    qui: mean EY1 EY0
    matrix ATT =  r(table)
    display "The ATT is: "  ATT[1,1] + ATT[1,2] // Applying the G-formula
    drop EY1 EY0
    // Also one can try
    gen ATT = ((y11[3,1]-y01[3,1]))*gendermatet + ((y10[3,1]-y00[3,1]))*genderfatet
    qui sum ATT
    drop ATT
\end{lstlisting}

\subsubsection{Statistical inference: The bootstrap}
\noindent Note that classical statistical inference methods do not apply for the causal inference estimators. This is because the standard errors (SE) of the causal inference estimators do not account for the different steps we need to make when we balance the confounders between exposure groups. We use the bootstrap procedure for inference implemented in Stata with the command \textit{bootstrap}.\citep{Efron1993AnBootstrap} Usually, the bootstrap is used to approximate the variance of the estimator (i.e., G-computation or IPTW for the ATE).\cite{Efron1993AnBootstrap, Efron1982ThePlans} When estimating the variance using the bootstrap method, the observed data is thought of as representing the entire target population, and each draw (with replacement) from data mimics the sampling variability. Under certain assumptions, this set of draws will return estimates of the sampling distribution that are equivalent to having actually repeated the sampling from the original target population.\citep{Efron1993AnBootstrap} Typically, for procedures that use parametric models, the bootstrap would give reliable estimation for the variance of the estimator. However, note, it does not account for the bias engendered by model misspecification, so it just provides sampling variability for whatever the estimator is estimating.\citep{Efron1993AnBootstrap} The accuracy with which the bootstrap distribution estimates the sampling distribution depends on the number of observations in the original sample and the number of replications in the bootstrap. To implement the bootstrap procedure in Stata we need to define a program that estimates the non-parametric G-formula and then samples (with replacement) the ATE to derive the confidence intervals for the ATE and the ATT. In box 6 we provide the code to compute the SE for the ATE, and the ATT, using Stata.
\\

\noindent\textbf{Box 5: Bootstrap 95\% Confidence Intervals (CI) for the ATE/ATT estimated using the Non-parametric G-Formula}
\begin{lstlisting}[escapeinside={(*}{*)}]
    // ATE
    capture drop program ATE
    program define ATE, rclass      // As before but now define a program to estimate the ATE
       capture drop y1
       capture drop y0
       capture drop ATE
       sumup $Y, by($A $C)
       matrix y00 = r(Stat1)
       matrix y01 = r(Stat2)
       matrix y10 = r(Stat3)
       matrix y11 = r(Stat4)
       gen ATE = ((y11[3,1]-y01[3,1]))*genderm + ((y10[3,1]-y00[3,1]))*genderf
       qui sum ATE
       return scalar ate = (*\color{blue}{\textasciigrave r(mean)\textquotesingle}*)
    end
    qui bootstrap r(ate), reps(1000): ATE   // Bootstrap 1000 estimates of the ATE 
    estat boot, all
    drop ATE
    
    // ATT
    capture program drop ATT
    program define ATT, rclass
       capture drop y1
       capture drop y0
       capture drop ATT
       sumup $Y, by($A $C)
       matrix y00 = r(Stat1)
       matrix y01 = r(Stat2)
       matrix y10 = r(Stat3)
       matrix y11 = r(Stat4)
       gen ATT = ((y11[3,1]-y01[3,1]))*gendermatet + ((y10[3,1]-y00[3,1]))*genderfatet
       qui sum ATT
       return scalar att = (*\color{blue}{\textasciigrave r(mean)\textquotesingle}*)
    end
    qui bootstrap r(att), reps(1000): ATT
    estat boot, all
    drop ATT
\end{lstlisting}

\noindent After bootstrapping, the estimate of the ATE and the ATT was 7.4\% and 7.3\%, respectively. We obtain three sets of confidence intervals for each of ATE and ATT. For the ATE, the first (N) is an approximation based on the normal distribution (95\% CI 4.79-9.95), the second (P) is based on the percentile of the bootstrap distribution (4.74-9.84), and the third (BC) is based on the bias-corrected (4.91 - 10.01) (Table 3). We now do the same for the ATT. The 95\% CIs for the ATT are: normal-approximation (4.74-9.91), percentile (4.71-10.00), and bias-corrected (4.70-10.00). Note that the percentile interval is a simple "first-order" interval that is formed from quantiles of the bootstrap distribution. However, it is based only on bootstrap samples and does not adjust for skewness in the bootstrap distribution. Thus, we will report the BC 95\% CI.\citep{Jung2019ComparisonSimulation}
\\

\noindent Remaining with the non-parametric implementation of the G-formula, we turn our attention to computing the ATE and ATT using a fully saturated regression model (still with only one confounder) using the full information of the sample (including the interactions). In Stata there are two different approaches that we illustrate in boxes 7 and 8 using the commands \textit{predictnl} and \textit{margins}.
\\

\noindent\textbf{Box 6: Non-parametric G-Formula using a fully saturated regression model in Stata (A)} \\
\noindent To estimate the ATE using a fully saturated regression model we need to include all the possible interactions between the treatment and the different levels of the confounders (if categorical otherwise with a continuous confounder). We use the \# facility of Stata to include the interaction between A and C. The \textit{coeflegend} option asks Stata to provide the list of the labels of the variables in the analysis. The labels are then used for the \textit{predictnl} command, which allows the computation of the non-parametric predictions based on the combination of the conditional probabilities from the regression coefficients. Finally, we average over the predictions to get the non-parametric estimate for the ATE but biased 95\% CI.  
\\
\begin{lstlisting}
    regress $Y ibn.$A ibn.$A#c.($C), noconstant vce(robust) coeflegend
    predictnl ATE = (_b[1bn.rhc] + _b[1bn.rhc#c.gender]*gender) - (_b[0bn.rhc] + _b[0bn.rhc#c.gender]*gender)
    mean ATE
\end{lstlisting}    

\noindent\textbf{Box 7: Non-parametric G-Formula using a fully saturated regression model in Stata (B)} \\
A simpler option would be to use the \textit{margins} command to estimate the marginal probabilities using the option \textit{vce(unconditional)}. Then the difference in marginal probabilities between the treated versus non-treated is implemented using the contrast option from the \textit{margins} command. Note, that here we obtain the same estimate of the ATE as 7.4\% (95\%CI 4.83-9.91) but the correct 95\%CI has been calculated using the Delta method (Table 3). The Delta method is a statistical approach to derive standard error of an asymptotically normally distributed estimator. It uses a first order Taylor approximation which is how we approximate the distribution of a function using a tangent line (i.e., the first derivative).\citep{Oehlert1992AMethod} Therefore, using the Delta method here we assume that the ATE estimate from the G-computation is normally distributed and asymptotically linear.\citep{Kennedy2016SemiparametricInference}
\\

\begin{lstlisting}
    regress $Y ibn.$A ibn.$A#c.($C), noconstant vce(robust) // Fully saturated model specification
    margins $A , vce(unconditional) // Marginal probability for A
    margins r.$A , contrast(nowald) // Difference in marginal probability between treatment groups
\end{lstlisting}

\subsection{Parametric G-formula}
\noindent In contrast to the non-parametric methods (i.e., probability distribution free), to compute the ATE parametrically we assume there is a particular probability distribution that fits the distribution of our data. To compute the ATE, we first regress (using a simple linear regression model) the outcome over the confounder separately for each treatment group, and then contrast the difference in the expected probabilities between the two treatment groups. The algebraic form of the ATE under the G-computation is given by

\begin{equation}
\text{ATE}=\frac{1}{n} \sum_{i=1}^{n} \left(\hat{E}(Y_{i}\mid A_{i}=1, \boldsymbol{W}_{i})\,-\,\hat{E}(Y_{i}\mid A_{i}=0, \boldsymbol{W}_{i})\right)
\end{equation}
\\

\noindent In box 8 we provide the code to compute, by hand, the ATE based on the parametric G-formula for one confounder using parametric regression adjustment. 
\\

\noindent \textbf{Box 8: Parametric regression adjustment implementation of the G-Formula}
\begin{lstlisting}
    regress $Y $C if $A==1       // Expected probability amongst those with RHC
    predict double y1hat
    regress $Y $C if $A==0       // Expected probability amongst those without RHC
    predict double y0hat
    mean y1hat y0hat             // Difference between the expected conditional probabilities
    lincom _b[y1hat] - _b[y0hat] //  ATE and biased confidence interval 
\end{lstlisting}

\noindent The risk of mortality amongst those with RHC is 7.4\% higher compared to those without RHC. Note that if we want to naively compute a 95\% for the linear contrast between the marginal potential outcomes we will get a biased CI that does not account for the two-step procedure to get the marginal probabilities.
\\

\noindent \textbf{Box 9: Parametric regression adjustment using Stata's \textit{teffects}}\\
\noindent In box 10, we just confirm the result we got for the ATE estimate using the STATA's \textit{teffects} command including the '\textit{ra}' option to perform the regression adjustment. Note the difference between the naive and the \textit{teffects} 95\% CIs. The \textit{teffects} uses the Delta method to correct for the two step estimation procedure and provide appropriate statistical inference.
\\

\begin{lstlisting}
    teffects ra ($Y $C) ($A) //Parametric G-Formula implementation in Stata
\end{lstlisting}

\noindent With the \textit{teffects} command in Stata the ATE is 7.4\%  which is the same as we obtained by hand. However, note that the 95\%CI for the ATE using the command from Stata is more conservative than using the naive approach without accounting for the two-step estimation procedure (i.e., 95\%CI: 4.83 - 9.91 and 95\%CI: 7.35 - 7.39, respectively for the \textit{teffects} and the naive approaches) (Table 3).
\\

\noindent \textbf{Box 10: Bootstrap for the parametric regression adjustment}\\
\noindent Again, if we want to compute the 95\% CI by hand using Stata we must use the bootstrap procedure (refer to Box 6 for an explanation). 
\\

\begin{lstlisting}[escapeinside={(*}{*)}]
    capture program drop ATE
    program define ATE, rclass
       capture drop y1
       capture drop y0
       reg $Y $C if $A==1 
       predict double y1, xb
       quiet sum y1
       reg $Y $C if $A==0 
       predict double y0, xb 
       quiet sum y0
       mean y1 y0 
       lincom _b[y1]-_b[y0]
       return scalar ace =(*\color{blue}{\textasciigrave r(estimate)\textquotesingle}*)
    end
    qui bootstrap r(ace), reps(1000): ATE
    estat boot, all
    drop ATE
\end{lstlisting}

\noindent After bootstrapping, the estimate of the ATE is 7.4\% and the bias-corrected 95\%CI: (4.73-10.23) (Table 3).
\\

\noindent \textbf{Box 11: Parametric multivariable regression adjustment implementation of the G-Formula}\\
\noindent As is often the case, there is almost always more than one confounder. The parametric computation of the G-formula can easily be extended to include more than one confounder: remember that \textbf{W} includes all of the confounders in our study. 
\\
\begin{lstlisting}
    regress $Y $W if $A==1          // Saturated regression model with all confounders for those with RHC
    predict double y1hat 
    regress $Y $W if $A==0          // Saturated regression model with all confounders for those without RHC
    predict double y0hat
    mean y1hat y0hat                // ATE is the difference in expectations
    lincom _b[y1hat] - _b[y0hat]    // The estimation from this command gives a biased confidence interval for the ATE
\end{lstlisting}

\noindent The ATE of those with RHC (i.e. risk of mortality amongst those with RHC) is 8.3\%, (95\% CI: 8.22-8.43) higher compared to those without RHC. Note that the 95\%CI provided by the \textit{lincom} Stata command is biased as it is not accounting for the two-step estimation procedure to derive the ATE.
\\

\noindent \textbf{Box 12: Parametric multivariable regression adjustment using Stata's \textit{teffects} command}\\
\noindent In box 13 we use the Stata's teffects command to confirm our results: note that we now include \textbf{W} instead of the single confounder \textbf{C}.
\\
\begin{lstlisting}
    teffects ra ($Y $W) ($A)
\end{lstlisting}

\noindent We obtain the same results with Stata's command as with our calculations by hand (ATE 8.3\%; 95\% CI: 5.80-10.85). However, note again the difference between the 95\% CI estimated naively and using the \textit{teffects} command (Table 3).
\\

\noindent \textbf{Box 13: Parametric multivariate regression adjustment using Stata's \textit{margins} command}\\

\noindent In box 14 we show another way of obtaining the ATE under the parametric G-formula approach using the Stata \textit{margins} command after fitting a fully saturated regression model. First, we regress the dependent variable (Y) over the treatment (A) and the interaction of A with all of the other independent variables (W). We do it using the Stata functionalities \textit{ibn} to force the estimation of all the levels of a categorical variable and \text{\#} to indicate the interaction between all the levels of A with all of the other variables included in the model (W). Then, the \textit{margins} command calculates the predicted value of the expectation of the outcome given the treatment and the confounders i.e. E(Y|A,\textbf{W}) and then reports the mean value of those predictions for each level of the treatment (A). The \textit{vce} option specified whether the standard errors should account for heteroscedasticity: \textit{robust} accounts for this but \textit{unconditional} does not. Finally, to compute the ATE and provide corrected 95\% CI based on the Delta method, we use the \textit{contrast} option to contrast the outcomes between those who received RHC and those who did not. Note the results are the same as before using the \textit{teffects} command (i.e., ATE 8.3\%; 95\% CI: 5.80-10.85) (Table 3).\\

\begin{lstlisting}
    regress $Y ibn.$A ibn.$A#c.($W), noconstant vce(robust) // Fully saturated model
    margins $A, vce(unconditional) // E(Y|A=1,W), E(Y|A=0,W) and Delta method for the standard errors (i.e., vce unconditional) and 95%CI 
    margins r.$A, contrast(nowald) // ATE and Delta method for the standard error and 95%CI 
\end{lstlisting}

\noindent \textbf{Box 14: Bootstrap for the multivariable parametric regression adjustment}\\
\noindent Finally, in box 15 we show how to compute the bootstrap 95\% CIs for the G-computation implementation of the G-formula using regression adjustment in Stata.
\\

\begin{lstlisting}[escapeinside={(*}{*)}]
    capture program drop ATE
    program define ATE, rclass
       capture drop y1
       capture drop y0
       reg $Y $W if $A==1 
       predict double y1, xb
       quiet sum y1
       reg $Y $W if $A==0 
       predict double y0, xb 
       quiet sum y0
       mean y1 y0 
       lincom _b[y1]-_b[y0]
       return scalar ace =(*\color{blue}{\textasciigrave r(estimate)\textquotesingle}*)
    end
    qui bootstrap r(ace), reps(1000): ATE dots
    estat boot, all  
    drop ATE
\end{lstlisting}

\noindent After bootstrapping, the estimate of the ATE is 8.3\%. The bootstrapped bias-corrected 95\%CI are (5.77-10.83) (Table 3).
\\

\noindent \textbf{Box 15: Computing the parametric marginal risk ratio after regression adjustment}\\
\noindent Before moving onto other methods, we show that \textit{teffects} can be used for  different estimands such as the marginal Risk Ratio (RR), which is the ratio between the E(E(Y|A=1,\textbf{W})) and E(E(Y|A=0,\textbf{W})). To do this in Stata we use the \textit{aequations} option to display the regression model coefficients used to predict the outcomes as well as the coefficients used to predict the exposure in order to manipulate them and compute the marginal RR estimate as follows: \\

\begin{equation*}
   100\times\,\left(\frac{E_{w}(\hat E(Y|A=1,\textbf{W})\,-\,(\hat E(Y|A=0,\textbf{W}))}{E_{w}(\hat E(Y|A=0,\textbf{W}))}\,-\,1 \right)\,=\, \frac{E_{w}(\hat E(Y|A=1,\textbf{W}))}{E_{w}(\hat E(Y|A=0,\textbf{W}))}
\end{equation*} \\

\noindent Note that again we use the Delta method to derive the 95\%CI for the marginal RR using the \textit{nlcom} command in Stata.\\

\begin{lstlisting}
	teffects  ra ($Y $W) ($A), aequations // aequations display the estimation auxiliary equations
	teffects ra ($Y $W) ($A), coeflegend // coeflegend displays the name of the coefficients
	nlcom  100*_b[ATE:r1vs0.$A]/_b[POmean:0.$A] // nlcom uses the Delta method to compute the standard error 
	// 27.4% increase in risk
	teffects  ra ($Y $W) ($A), pom coeflegend // pom displays E(Y|A=1,W) and E(Y|A=0,W), respectively 
	nlcom _b[POmeans:1.rhc]/ _b[POmeans:0bn.rhc]
	// 27.4% increase  in  relative  risk
\end{lstlisting}

\noindent The relative risk of mortality is 27\% (95\% CI 18.25 - 36.61) higher amongst those with RHC compared to those without RHC. Note that using the Stata command \textit{nlcom} we estimate the SE for the RR based on the Delta method accounting for the two-step estimating procedure. 
\\

\section{Inverse probability of treatment weighting}

\subsection{Inverse probability weighting based on the propensity score}

In observational settings, some individuals will be more likely than others to be exposed (A) due to their characteristics. Suppose some individuals who were exposed were unlikely to be exposed based on a specific set of features encapsulated in a particular vector of confounders (W). To balance the exposure risk we increase the weight (upweight) these individuals on the outcome variable (Y) by the inverse of their probability of treatment or exposure (A) (i.e., propensity score), so that they represent themselves but also the other individuals with similar characteristics, who were unexposed. Concurrently, we upweight those patients that were unlikely to be unexposed. The resulting dataset is unchanged apart from now A and \textbf{W} are independent. Therefore, a comparison of $Y(1)$ to $Y(0)$ gives a marginal causal effect under the three identification assumptions and further assuming the propensity score model is correctly specified. Therefore, the inverse probability of treatment weighting (IPTW) estimators also compute the G-formula parametrically but based on the inverse probability of treatment or exposure.\citep{Robins1986AEffect, Rosenbaum1983TheEffects} Originally, they were motivated from the classical Horvitz and Thompson survey estimator used to upweight on the outcome variable by the inverse probability that it is observed to account for the missingness process.\citep{Horvitz1952AUniverse} Therefore, regression adjustment and IPTW estimators are both computational implementations of the G-formula, i.e. generalisation of the standardisation procedure. Thus, these methods are also known as G-computation methods as they are based on the computational implementation of the G-Formula.
\\

\noindent Analytically departing from the identification assumptions for the ATE for the regression adjustment G-computation estimand (ATE = $E_{w}$(E(Y|A=1,\textbf{W}) - $E_{w}$(Y|A=0,\textbf{W})), we can rewrite the same estimand as a function of the distribution of A given W i.e., P(A|\textbf{W}) a.k.a propensity score or treatment mechanism.\\

\noindent Therefore, the IPTW estimand of the ATE is given by

\begin{equation}
\text{ATE}\,=\,\hat{E}\left(\frac{\mathbb{I}(A = a)}{g(A = 1 \mid \boldsymbol{W})}\,Y\right),
\end{equation}
\\
\noindent where $g(A=a\mid \boldsymbol{W}) = P(A=a\mid \boldsymbol{W})$ and $ \mathbb{I} $ refers to the binary indicator A = a set equal to 0 or 1.

\noindent There is a modified version of the IPTW estimator in (5) that is more efficient and is used in practice, consisting on a stabilised version of the weights (i.e., ranging between zero and one). The stabilised version of the IPTW estimator is given by

\begin{equation}
\text{ATE} \,=\,\frac{\hat E\left(\frac{\mathbb{I}(A=a)}{g(A=1\mid\boldsymbol{W})}\,Y\right)}{\hat E \left(\frac{\mathbb{I}(A=a)}{g(A=1\mid\boldsymbol{W})}\right)},
\end{equation}
\\
\noindent where $g(A=a\mid \boldsymbol{W}) = P(A=a\mid \boldsymbol{W})$ and $ \mathbb{I} $ refers to the binary indicator A = a set equal to 0 or 1.

\noindent By repeated use of the law of total expectation, the IPTW and the G-computation regression adjustment estimands for a binary treatment are equivalent as given by

\begin{equation*} 
\begin{split}
E\left(\frac{(\mathbb{I}(A=1)}{P(A =1 \mid \boldsymbol{W})}\,Y\right)\,
&=\,E \left( E \left( \frac{\mathbb{I}(A = 1)\,Y}{P(A = 1\mid \boldsymbol{W})}\bigl\lvert A,\boldsymbol{W}\right)\right) \\
&=\,E \left( \frac{\mathbb{I}(A = 1)}{P(A = 1\mid \boldsymbol{W})} E(Y \mid A, \boldsymbol{W})\right) \\
&=\,E \left( \frac{\mathbb{I}(A = 1)}{P(A = 1\mid \boldsymbol{W})} E(Y \mid A = 1,\boldsymbol{W})\right) \\
&=\,E \left(E\left( \frac{\mathbb{I}(A = 1)}{P(A = 1\mid \boldsymbol{W})}E(Y \mid A = 1, \boldsymbol{W}) \bigl\lvert \boldsymbol{W} \right)\right) \\
&=\,E \left(\frac{E(Y \mid A, \boldsymbol{W})}{P(A = 1\mid \boldsymbol{W})} P(A = 1\mid \boldsymbol{W})\right)\,=\, E[E(Y \mid A = 1, \boldsymbol{W})].
\end{split}
\end{equation*}
\\

\noindent In box 16 we show how to compute the IPTW by hand in two steps: 
\begin{itemize}
    \item First, the propensity score model is fitted (i.e., a logistic regression model for a binary treatment; however, data-adaptive and machine learning techniques can be used too)
    \item Then the sampling weights are generated based on the inverse probability of treatment. Note that the weights are just the implementation of the classical Horvitz-Thompson survey estimator,\citep{Horvitz1952AUniverse} (see rows 3 and 4) also known as unstabilised weights.
 \end{itemize}   
Note that when there are near violations of the positivity assumption the unstabilised weights can have large values increasing the variance and hence the uncertainty of the ATE estimation. Therefore, it is advisable to explore the distribution of the weights in order to evaluate the extent to which they balance the distribution of confounders across the levels of the treatment (i.e.. equally distributed). It is common to provide a table with the standardised treatment differences between unweighted and weighted confounders. The visualization of the overlap of the propensity scores by level of the treatment is commonly used to identify and visualise positivity or near positivity violations.  
\\

\noindent \textbf{Box 16: Computation of the IPTW estimator for the ATE}\\
\begin{lstlisting}
    logit $A $W, vce(robust) nolog          // Propensity score model for the exposure
    predict double ps                       // Propensity score prediction
    generate double ipw1 = ($A==1)/ps       // Sampling weights for the treated group
    generate double ipw0 = ($A==0)/(1-ps)   // Sampling weights for the non-treated group
    regress $Y [pw=ipw1]                    // Weighted outcome probability among treated
    scalar Y1 = _b[_cons]
    regress $Y [pw=ipw0]                    // Weighted outcome probability among non treated
    scalar Y0 = _b[_cons]
    display "ATE =" Y1 - Y0
\end{lstlisting}
\noindent You will notice that the risk difference between those with RHC and those without is 8.3\% (confidence intervals are estimated below in box 18).
\\

\noindent \textbf{Box 17: Bootstrap computation for the IPTW estimator}\\
\noindent As before, we can obtain confidence intervals using the bootstrap procedure implemeted in box 18.
\\
\begin{lstlisting}[escapeinside={(*}{*)}]
    program drop ATE
    program define ATE, rclass
      capture drop y1
      capture drop y0
      regress $Y [pw=ipw1]
      matrix y1 = e(b)
      gen double y1 = y1[1,1]
      regress $Y [pw=ipw0]
      matrix y0 = e(b)
      gen double y0 = y0[1,1]
      mean y1 y0
      lincom _b[y1]-_b[y0]
      return scalar ace = (*\color{blue}{\textasciigrave r(estimate)\textquotesingle}*)
    end
    qui bootstrap r(ace), reps(1000): ATE
    estat boot, all   
    drop ATE
\end{lstlisting}
\noindent After bootstrapping, the estimate of the ATE is 8.3\%. The bootstrapped bias-corrected confidence are: (5.71-10.90).
\\

\noindent \textbf{Box 18: Computation of the IPTW estimator for the ATE using Stata's \textit{teffects} command}\\
\noindent We now confirm this result in box 19 using Stata's \textit{teffects} command. Note that the Horvitz-Thompson estimator is implemented using the \textit{ipw} option. We obtain the same point estimate for the ATE and slightly different, but consistent, 95\%CI based on the robust SE derived from the functional Delta method (i.e., ATE 8.3\%; 95\% CI 5.78 - 10.83) (Table 2).
\\
\begin{lstlisting}
    teffects ipw ($Y) ($A $W, logit), nolog vsquish
\end{lstlisting}

\noindent \textbf{Box 19: Assessing IPTW balance}\\
\noindent In box 19 we show how to explore "balance" i.e., that the distribution of the confounders are balanced between those with RHC and those without, after re-weighting the contributions of patients using IPTW. When applying weights we must be careful as we are assuming that the exposure has been balanced across the levels of the confounders. In Stata we use the \textit{tebalance} option after using the \textit{teffects} command but the balance can be assessed by hand as well.
\\

\begin{lstlisting}
    qui teffects ipw ($Y) ($A $W)   
    tebalance summarize // Stata's tebalance 
    
    // tebalance by hand (gender)
    egen genderst = std(gender) // Standardization
    logistic $A $W // Propensity score
    predict double ps
    gen ipw = .
    replace ipw=($A==1)/ps if $A==1
    replace ipw=($A==0)/(1-ps) if $A==0
    regress genderst $A // Raw difference
    regress genderst $A [pw=ipw] // Standardized difference
\end{lstlisting} 

\noindent After weighting, the two exposure groups appear to be well balanced. Prior to weighting, there was some imbalance between genders, education level and those who had a carcinoma (absolute values of the standardised differences close to or beyond 10\%). A variance ratio equal to 1 before and after weighting informs us that the distribution of the exposure across the levels of the confounder is the same (i.e., perfectly balanced). Note that the variance ratio for the continuous variable age is 0.79, which is a bit further from 1 than the variance ratio for the original (unweighted) sample (Table (\ref{table:2})). 

\begin{center}
    \begin{table}[ht!]
    \centering
    \noindent \caption{Distribution of the confounders before and after applying weights}
    \label{table:2}
    \begin{threeparttable}
	\begin{tabular}{ l r r r r r } \\
            & \multicolumn{2}{c}{\textbf{Standardised differences}} & &                             \multicolumn{2}{c}{\textbf{Variance ratio}} \\ [0.75ex] \cline{2-3} \cline{5-6} 
\textbf{Confounder} & \textbf{Raw} & \textbf{Weighted} & & \textbf{Raw} & \textbf{Weighted} \\
\hline \\
    Gender      &  0.0931272 &  0.0004124 & & 0.9771947 & 0.9999057 \\
    Age         & -0.0613524 & -0.0038196 & & 0.8174922 & 0.7899075 \\
    Education   &  0.0913642 & -0.0025822 & & 1.0147230 & 1.0250380  \\
    Race        & -0.0022396 &  0.0023428 & & 1.0295870 & 1.0254230  \\
    Carcinoma   & -0.1051837 &  0.0012232 & & 0.8386081 & 1.0226510  \\
    \hline
    \end{tabular}
    \end{threeparttable}
    \end{table}
\end{center}

\noindent There is no definitive value at which the exposure is considered unbalanced; however, as a guideline, a variance ratio less than 0.5 must clearly indicate that the data is not balanced and the potential for the positivity violation must be explored (i.e., when $P(A=a\mid C=c)$ is near to zero or one). An additional strategy is to check the distribution of the weights: if very large weights are identified this confirms the violation of the positivity assumption. Again there is no clear consensus but when there are very large weights researchers often set the weights to a less extreme value. This is done by truncating the distribution of the confounder to the $5^{th}$ and $95^{th}$ percentiles: removing the data at the extremes of the distribution. Note that "truncating" or "trimming" the weights reduces non positivity and variance but at the expense of introducing bias. \citep{Cole2008ConstructingModels}. In extreme cases, changing the estimand to the ATT could be another solution.
\\

\begin{figure}[ht!]
\begin{center} 
\caption{Propensity score overlap by exposure status}{\includegraphics[scale=0.85]{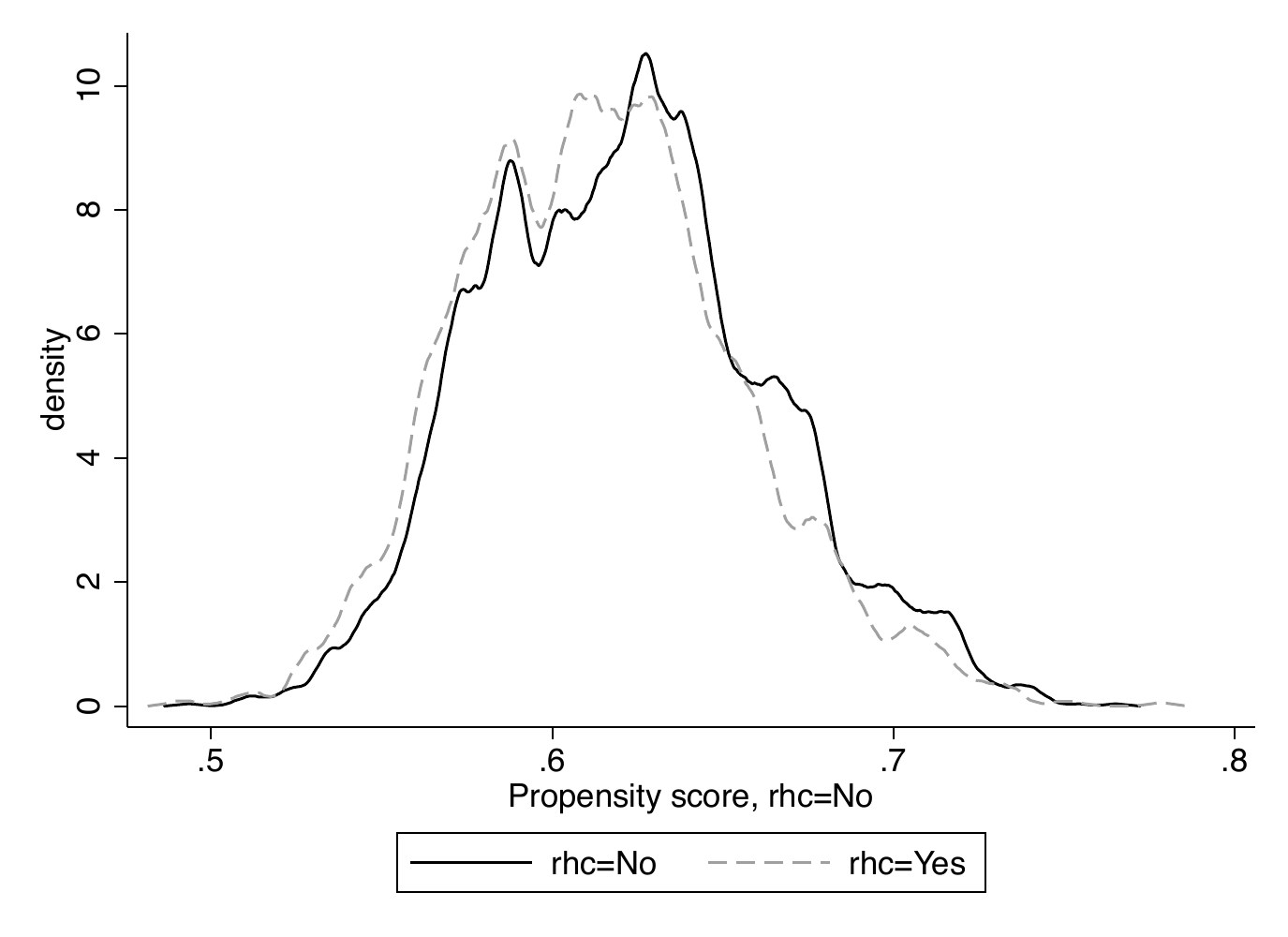}\hfill}\label{Figure:2}
\end{center}
\end{figure}

\noindent \textbf{Box 20: Assessing IPTW overlap by hand}\\
\noindent Therefore, independently of the balance, it is also important to check for any violations of the identifiability conditions. Starting with the positivity assumption, overlapping the weights gives a visual impression of whether the weights are balanced. In box 21 we show how to visualise the "overlap" using a kernel density estimate by the levels of the exposure. (Figure (\ref{Figure:2})). shows that there appears to be a suitable amount of overlap. 
\\

\begin{lstlisting}
    sort $A
    by $A: summarize ps
    kdensity ps if $A==1, generate(x1pointsa d1A) nograph n(10000) // Non-parametric kernel density estimate of the distribution of the propensity score among treated patients
    kdensity ps if $A==0, generate(x0pointsa d0A) nograph n(10000) // Non-parametric kernel density estimate of the distribution of the propensity score among non-treated patients
    label variable d1A "density for RHC=1"
    label variable d0A "density for RHC=0"
    twoway (line d0A x0pointsa , yaxis(1))(line d1A x1pointsa, yaxis(2))
\end{lstlisting}

\noindent \textbf{Box 21: Assessing overlap using \text{teffects overlap}}\\
\noindent Note that the overlap plots can be obtained with Stata's command \textit{overlap} after calling the \textit{teffects} command. Note that we are using data where the balance and overlap are nearly perfect but in real-world observational data the balance and overlap will differ.
\\
\begin{lstlisting}
    qui: teffects ipw ($Y) ($A $W, logit), nolog vsquish
    teffects overlap    
\end{lstlisting}

\subsection{Marginal structural model with stabilised weights}

A marginal structural model (MSM) is a weighted regression model for the outcome (Y) as dependent variable and treatment or exposure (A) as independent variable. In box 23 we show how to compute a MSM. Note that the coefficient of the MSM for the treatment A is the ATE estimate. To fit the MSM we use as sampling weights (i.e., Horvitz and Thompson) the IPTW computed as the inverse of the probability of treatment or propensity score (see box 17). However, the variance of the ATE estimate can be inflated in a non-saturated MSM due to the presence of large weights. Larger weights are the result of potential near-violation of the positivity assumption. Thus, to counter it, one may use a stabilised version of the weights.\citep{Robins2000MarginalEpidemiology} Stabilised weights only differ from the weights described above in that they do not take a simple inverse but instead divide the baseline probability of selecting a treatment (estimated from a model with no covariates) by the probability of selecting treatment given the covariates. The mean of the stabilised weights is equal to one. Therefore, the stabilised weights produce ATE estimates that have smaller variance being more robust against near-positivity violations. Finally, note that for statistical inference we use the \textit{vce(robust)} to estimate the correct SE for the ATE using the Delta method.\citep{Boos2013EssentialMethods}
\\

\noindent \textbf{Box 22: Computation of the IPTW estimator for the ATE using a MSM}
\begin{lstlisting}
  // baseline treatment probabilities
  logit $A, vce(robust) nolog
  predict double nps, pr
  // propensity score model
 logit $A $W, vce(robust) nolog
  predict double dps, pr
  // Unstabilised weight
  gen ipw = .
  replace ipw=($A==1)/dps if $A==1
  replace ipw=($A==0)/(1-dps) if $A==0
  sum ipw 
	
  // Stabilized weight 
  gen sws = .
  replace sws = nps/dps if $A==1
  replace sws = (1-nps)/(1-dps) if $A==0
  sum sws
    
  // MSM 
  reg $Y $A [pw=ipw], vce(robust) // MSM unstabilized weight
  reg $Y $A [pw=sws], vce(robust) // MSM stabilized weight
\end{lstlisting}

\subsection{Inverse probability weighting plus regression adjustment}

\noindent The IPTW-RA is a estimator using a G-computation regression adjustment that incorporates the estimated stabilised IPTW. It has been shown that the IPTW-RA helps to correct the estimator when the regression function is misspecified provided that the propensity score model for the treatment is correctly specified. When the regression function is correctly specified, the weights do not affect the consistency of the estimator even if the model from which they are derived is missepecified.\citep{Kang2007DemystifyingData} Note that by blending both approaches we only increase our chances to get a more consistent estimate of the ATE and this is why estimators that combine both modelling approaches are named double robust. When one uses G-computation methods only, they simply rely on extrapolation of the treatment effects when there are identifiability issues due to sparsity and near-positivity violations. Adding the IPTW to the regression adjustment allows evaluation of the balance of the treatment and of possible positivity violations, increasing the researcher's awareness of the limitations of causal inference modelling. We also encourage researchers when possible to explore the implementation of the non-parametric G-formula (just using some of the most important confounders) to really understand their data at hand and identify potential problems relative to the curse of dimensionality from finite samples (i.e., zero empty cells for a given combination of conditional probabilities from the different variables included in analysis needed to implement the G-formula). 
\\

\noindent Although IPTW with regression adjustment (IPTW-RA) is usually more efficient than IPTW, it also relies on different parametric modelling assumptions: i) a parametric G-computation regression adjustment model, and ii) a parametric logit model for the propensity score of binary exposures. The G-computation weighted model uses the weights calculated from the predictions of the propensity score logit model. An estimated propensity score that is close to 0 or 1 is problematic, since it implies that some individuals will receive a very large weight leading to imprecise and unstable estimates. Here, the positivity assumption is required (i.e., $\text{if } P(C=c) > 0 \text{ then } P(A=a\mid C=c) > 0 \; \forall \; C \in c, a \in \{0, 1\}$) and stabilised weights are used (see code from Box 23). After reweighting the individuals, we generate a new pseudo-population (weighted population) from which the data generation does not follow a theoretical distribution and individuals are no longer independent. Therefore, we must use robust standard errors based on the Delta method applied to a function such as the Sandwich estimator of the SE.\citep{TsiatisSemiparametricExtranjeros}
\\

\noindent \textbf{Box 23: Computation of the IPTW-RA estimator for the ATE and bootstrap for statistical inference}
\\
\begin{lstlisting}[escapeinside={(*}{*)}]
    capture program drop ATE
    program define ATE, rclass
      capture drop y1
      capture drop y0
      reg $Y $W if $A==1 [pw=sws]
      predict double y1, xb
      quiet sum y1
      return scalar y1=(*\color{blue}{\textasciigrave r(mean)\textquotesingle}*)
      reg $Y $W if $A==0 [pw=sws]
      predict double y0, xb 
      quiet sum y0
      return scalar y0=(*\color{blue}{\textasciigrave r(mean)\textquotesingle}*)
      mean y1 y0 
      lincom _b[y1]-_b[y0]
      return scalar ate =(*\color{blue}{\textasciigrave r(estimate)\textquotesingle}*)
    end
    qui bootstrap r(ate), reps(1000): ATE
    estat boot, all   
    drop ATE
\end{lstlisting}

\noindent After bootstrapping, the estimate of the IPTW-RA ATE is 8.3\%, bias-corrected 95\%CI (5.76-10.87). The results are the same as those obtained using the Stata's \textit{teffects} command with the option \textit{ipwra} presented in box 25 (i.e.,  ATE: 8.3\% and 95\% CI: 5.79-10.84) (Table 3).
\\

\noindent \textbf{Box 24: Computation of the IPTW-RA estimator for the ATE using Stata's \textit{teffects}}\\
\noindent Note that using \textit{ipwra} we specify two models (i.e., the model for the outcome and the model for the exposure). As with the parametric G-formula approach, we can also estimate the RR: 27.4\% higher mortality among treated than non-treated (95\% CI: 18.20-36.56), which is similar to the estimate obtained by the parametric G-formula.
\\

\begin{lstlisting}
	teffects  ipwra ($Y $W) ($A $W), pom coeflegend
	nlcom _b[POmeans:1.rhc]/ _b[POmeans:0bn.rhc] // marginal RR and 95%CI (Delta method)
	// 27.4% increased risk
\end{lstlisting}

\section{Augmented inverse probability weighting}

\noindent The AIPTW estimator is an IPTW estimator that includes an augmentation term, which corrects the estimator when the treatment model is misspecified. When the treatment model is correctly specified, the augmentation term vanishes as the sample size becomes large. Thus, the AIPTW estimator is more efficient than the IPTW. However, like the IPTW, the AIPTW does not perform well when the predicted treatment probabilities are too close to zero or one (i.e., near positivity violations). The augmentation term has expectation zero and includes the expectation of the propensity score and the regression adjustment outcome. Thus, the AIPTW integrates two parametric models i.e., a model for the outcome and a model for the treatment.\cite{Robins1994EstimationObserved, Bang2005DoublyModels} The AIPTW estimator produces a consistent doubly-robust estimate of the parameter if either of the two models has been correctly specified.\cite{Kang2007DemystifyingData, Bang2005DoublyModels} 
\\

\noindent Focusing on the IPTW estimator for the ATE in (3), let $\hat{\mu}$ be the expectation of the ATE using IPTW, more formally this is
\\

\begin{equation*}
    \hat{\mu_{a}} = \hat E\left(\frac{\mathbb{I}(A=a)}{g(A\mid \boldsymbol{W})}\,Y\right),
\end{equation*}
\\

\noindent we can rewrite the equation in the form of an estimating equation (see glossary) as
\\ 

\begin{equation*}
    \sum_{i}\left(\frac{A_{i}Y_{i}}{g(A_{i}\mid \boldsymbol{W}_{i})}-\mu_{a}\right) = 0,
\end{equation*}
\\

\noindent As long as the estimating function has mean zero then $\hat{\mu}$ is a consistent estimator of $\mu$, where $\hat{\mu_{a}}$ = $\hat E(Y_{i} \mid A_{i} = a, \boldsymbol{W_{i}}$). If we augment the estimating function using a mean-zero term, $\frac{A_{i}\,-\,g(A_{i}\mid \boldsymbol{W}_{i})}{g(A_{i}\mid \boldsymbol{W}_{i})}$, including the propensity score expectation ($g(A_{i}\mid \boldsymbol{W}_{i})$), we have integrated both the estimation of the treatment mechanism and the mean outcome ($E(Y_{i} \mid A_{i},\boldsymbol{W}_{i})$), then
\\

\begin{equation*}
    \sum_{i}\left(\frac{A_{i}Y_{i}}{g(A_{i}\mid \boldsymbol{W}_{i})} \,-\, \left( \frac{A_{i}\,-\,g(A_{i}\mid W_{i})}{g(A_{i}\mid \boldsymbol{W}_{i})} \right) E(Y_{i} \mid A_{i} = a, \boldsymbol{W_{i}}) \right) = 0,
\end{equation*}
\\

\noindent rearranging the equation we can see that the AIPTW estimator is a combination of inverse weighting and outcome regression defined for a binary treatment as 
\\

\begin{equation} 
\frac{1}{n}\underbrace{\sum_{i=1}^{n} \left((\hat Y_{i}\mid A_{i}=1, \boldsymbol{W}_{i}) \, - \, (\hat Y_{i} \mid A_{i}=0, \boldsymbol{W}_{i}) \right)}_{G-computation-Regression-Adjustment}
    \,+\,\frac{1}{n}\sum_{i=1}^{n}\underbrace{\left(\frac{A_{i}[Y_{i}-E(Y_{i} \mid A_{i}=1,\boldsymbol{W_{i}})]}{g(A_{i}=1 \mid \boldsymbol{W}_{i})}-\frac{(1-A_{i})[Y_{i}-E(Y_{i} \mid A_{i}=1,\boldsymbol{W_{i}})]}{g(A_{i}=0\mid \boldsymbol{W}_{i})}\right),}_{Zero-expectation}
\end{equation}
\\
\noindent where the ATE from the AIPTW estimator is defined as
\\
\begin{equation} 
\begin{split}
\text{AIPTW-ATE} \,=\,\mu_{1}\,-\,\mu_{0} \\
\mu_{1}\,&=\,\frac{1}{n}\sum_{i=1}^{n} \left((\hat Y_{i}\mid A_{i}=1, \boldsymbol{W}_{i})\,+\,\frac{A_{i}[Y_{i}-E(Y_{i} \mid A_{i}=1,\boldsymbol{W_{i}})]}{g(A_{i}=1 \mid \boldsymbol{W}_{i})}\right),
\\
\mu_{0}\,&=\,\frac{1}{n}\sum_{i=1}^{n} \left((\hat Y_{i} \mid A_{i}=0, \boldsymbol{W}_{i})
    \,+\,\frac{(1-A_{i})[Y_{i}-E(Y_{i} \mid A_{i}=1,\boldsymbol{W_{i}})]}{g(A_{i}=0\mid \boldsymbol{W}_{i})}\right).
\end{split}
\end{equation}
\\

\noindent Note that the second term in equation (7) can be interpreted as playing the role of two nuisance parameters of the AIPTW estimating function. The nuisance parameters are represented as a weighted sum of the residuals for the conditional mean of the outcome.\citep{Kennedy2016SemiparametricInference} The AIPTW equation (7) shows that the AIPTW estimator equals the G-formula estimator if the outcome model is correctly specified irrespective of the exposure model. Likewise, it will be equal to the IPTW estimator if the exposure model is correctly specified, irrespective of the outcome model.
\\

\noindent In box 25 we show how to compute the AIPTW estimator for the ATE using Stata.  
\\

\noindent \textbf{Box 25: Computation of the AIPTW estimator for the ATE and bootstrap for statistical inference}
\begin{lstlisting}[escapeinside={(*}{*)}]
		// Step (i) prediction model for the outcome using G-computation regression adjustment
		qui glm $Y $A $W, fam(bin) 
		predict double QAW, mu
		qui glm $Y $W if $A==1, fam(bin) 
		predict double Q1W, mu
		qui glm $Y $W if $A==0, fam(bin)
		predict double Q0W, mu
		
		// Step (ii): prediction model for the treatment
		qui logit $A $W
		predict double dps, pr
		qui logit $A
		predict double nps, pr
		gen sws = .
		replace sws = nps/dps if $A==1
		replace sws = (1-nps)/(1-dps) if $A==0
		
		// Step (iii): Estimation equation based on analytical formula 5
		gen double y1 = (sws*($Y-QAW) + (Q1W))
		quiet sum y1
		return scalar y1=(*\color{blue}{\textasciigrave r(mean)\textquotesingle}*)
		gen double y0 = (sws*($Y-QAW) + (Q0W))
		quiet sum y0
		return scalar y0=(*\color{blue}{\textasciigrave r(mean)\textquotesingle}*)
		mean y1 y0
		lincom _b[y1] - _b[y0]

		// step (iv) Bootstrap confidence intervals
		capture program drop ATE
		program define ATE, rclass
		capture drop y1
		capture drop y0
		capture drop Q*
		qui glm $Y $A $W, fam(bin) 
		predict double QAW, mu
		qui glm $Y $W if $A==1, fam(bin) 
		predict double Q1W, mu
		qui glm $Y $W if $A==0, fam(bin)
		predict double Q0W, mu
		gen double y1 = (sws*($Y-QAW) + (Q1W))
		quiet sum y1
		return scalar y1=(*\color{blue}{\textasciigrave r(mean)\textquotesingle}*)
		gen double y0 = (sws*($Y-QAW) + (Q0W))
		quiet sum y0
		return scalar y0=(*\color{blue}{\textasciigrave r(mean)\textquotesingle}*)
		mean y1 y0
		lincom _b[y1] - _b[y0]
		return scalar ate =(*\color{blue}{\textasciigrave r(estimate)\textquotesingle}*)
		end
		qui bootstrap r(ate), reps(1000): ATE
		estat boot, all
		drop ATE
\end{lstlisting}

\noindent After bootstrapping the ATE is 8.3\% and the bias-corrected 95\%CI confidence intervals are: (5.91-10.86) (Table 3). In box 27, we show the same results using Stata's \textit{teffects} command with the \textit{aipw} option. Note that we have to specify the model for the treatment and the model for the outcome. Likewise, we can compute the marginal RR as described above.
\\

\noindent \textbf{Box 26: Computation of the AIPTW estimator for the ATE and marginal risk ratio using Stata's \textit{teffects}}
\begin{lstlisting}
	teffects aipw ($Y $W) ($A $W, logit)
	// marginal Relative  Risk
	nlcom 100*_b[r1vs0.$A]/_b[POmean:0.$A]	
	//another way to compute it
	teffects aipw ($Y $W) ($A $W, logit), pom coeflegend
	nlcom _b[POmeans:1.rhc]/ _b[POmeans:0bn.rhc]
\end{lstlisting}
\noindent The ATE is 8.3\% (95\% CI: 5.79 - 10.84) (Table 3) and the marginal RR is 27.4\% (95\%CI: 18.2-36.6).
\\

\section{Data-adaptive estimation: Ensemble Learning Targeted Maximum Likelihood Estimation}

Targeted Maximum Likelihood Estimation (TMLE) is a semi-parametric doubly-robust method that reduces the bias of an initial estimate by targeting an estimate closer to the true model specification by allowing for flexible estimation using non-parametric data-adaptive machine-learning methods.\citep{vanderLaan2011TargetedData} It therefore requires weaker assumptions than its competitors. There are several tutorials published elsewhere but we recommend our tutorial as it provides a step-by-step guided implementation of TMLE illustrated in a realistic scenario based on cancer epidemiology where assumptions about correct model specification and positivity (i.e., when a study participant had zero probability of receiving the treatment) are nearly violated.\citep{Luque-Fernandez2018TargetedTutorial} The advantages of TMLE have repeatedly been demonstrated in both simulation studies and applied analyses.\citep{vanderLaan2011TargetedData} Evidence shows that TMLE consistently provides the least biased ATE estimate compared with other double-robust estimators such as the IPTW-RA and AIPTW.\citep{vanderLaan2011TargetedData}\\

\noindent In box 27, we provide the computational implementation of TMLE by hand (without data-adaptive estimation) to guide and interpret the different steps involved in the TMLE. A description of the theory behind these steps can be found elsewhere.\citep{Luque-Fernandez2018TargetedTutorial} In Step 1, we estimate the expected outcome (given treatment and confounders): this is called the plug in initial estimate of the estimator obtained via G-computation, namely $Q_{0}$. In Step 2, we define the expected exposure  given the confounders as we did previously for the estimation of the propensity score in box 17, namely $g_{0}$. Steps 1 and 2 are similar to the double-robust methods of AIPTW; however, we now come to the advantage of TMLE. In Step 3, we regress the predicted treatment values and predicted outcome introduced in the model as an offset on the observed outcome. The parameter estimates (epsilon) for that regression are used to correct the initial estimations of $Q_{0}$. In other words, we reduce the residual bias and optimized the bias-variance trade-off for the estimate of the ATE so that we can obtain valid statistical inference. Note that the TMLE framework adds the possibility to estimate the $Q_{0}$ and $g_{0}$ models using data adaptive machine learning algorithms and selecting the best model or an ensemble of the models.\citep{vanderLaan2011TargetedData} It has been shown that using machine learning algorithms reduces misspecification estimation bias.\citep{Naimi2017ChallengesAlgorithms}
\\

\noindent \textbf{Box 27: Computational implementation of TMLE by hand}
\begin{lstlisting}
* Step 1: prediction model for the outcome Q0 (g-computation)
glm $Y $A $W, fam(binomial)
predict double QAW_0, mu
gen aa=$A
replace $A = 0
predict double Q0W_0, mu
replace $A= 1
predict double Q1W_0, mu
replace $A = aa
drop aa

// Q to logit scale
gen logQAW = log(QAW / (1 - QAW))
gen logQ1W = log(Q1W / (1 - Q1W))
gen logQ0W = log(Q0W / (1 - Q0W))

* Step 2: prediction model for the treatment g0 (IPTW)
glm $A $W, fam(binomial)
predict gw, mu
gen double H1W = $A  / gw
gen double H0W = (1 - $A ) / (1 - gw)

* Step 3: Computing the clever covariate H(A,W) and estimating the parameter (epsilon) (MLE)
glm $Y H1W H0W, fam(binomial) offset(logQAW) noconstant
mat a = e(b)
gen eps1 = a[1,1]
gen eps2 = a[1,2]

* Step 4: update from Q0W and Q1W to Q0W_1 and Q1W_1
gen double Q1W_1 = exp(eps1 / gw + logQ1W) / (1 + exp(eps1 / gw + logQ1W))
gen double Q0W_1 = exp(eps2 / (1 - gw) + logQ0W) / (1 + exp(eps2 / (1 - gw) + logQ0W))

* Step 5: Targeted estimate of the ATE 
gen ATE = (Q1W_1 - Q0W_1)
summ ATE
global ATE = r(mean)

* Step 6: Statistical inference (functional Delta method)
qui sum(Q1W_1)
gen EY1tmle = r(mean)
qui sum(Q0W_1)
gen EY0tmle = r(mean)

gen d1 = (($A  * ($Y - Q1W_1)/gw)) + Q1W_1 - EY1tmle
gen d0 = ((1 - $A ) * ($Y - Q0W_1)/(1 - gw))  + Q0W_1 - EY0tmle

gen IC = d1 - d0
qui sum IC
gen varIC = r(Var) / r(N)

global LCI =  $ATE - 1.96*sqrt(varIC)
global UCI =  $ATE + 1.96*sqrt(varIC)
display "ATE:"  %05.4f  $ATE _col(15) "95%CI: " %05.4f  $LCI "," %05.4f  $UCI
\end{lstlisting}

\noindent Note, that in box 27, the residual bias is reduced by solving an equation that calculates how much to update, or fluctuate, our initial outcome estimates i.e., $$logit(E[Y|A,\boldsymbol{W}]) = logit(\hat{E}[Y|A,\boldsymbol{W}]) + \epsilon H(A,\boldsymbol{W}),$$ where E(Y|A,\textbf{W}) represents the expectation of the outcome (Y) given the treatment status (A) and the set of confounders (\textbf{W}). To solve this equation, we fit an intercept-free logistic regression (using the clever covariate \textit{H} as the only predictor of the observed outcome) and the initially predicted outcome (under the observed treatment) as an offset (step 3 rows 24-27). Fitting the logistic regression, using maximum likelihood procedures, solves an estimating equation which yields many useful statistical properties of TMLE, such as: 1) the final estimate is consistent as long as either the outcome or treatment model are estimated correctly (consistently); 2) if both of these models are estimated consistently, the final estimate achieves its smallest possible variance as the sample size approaches infinity (namely, efficiency). 
\\

\noindent Next, we added the coefficient $\epsilon$ of the clever covariate \textit{H} in the previous step to the expected outcome for all observations from the model fitted in Step 1 using (step 4: rows 30-31), updating the $Q_{0}$ and $g_{0}$ models predictions.
\\

$$\widehat{E}[Y|A=1,\boldsymbol{W}]) = expit(logit(E[Y|A=1,\boldsymbol{W}]) + \epsilon H(1,\boldsymbol{W})), \text{and}$$

$$\widehat{E}[Y|A=0,\boldsymbol{W}]) = expit(logit(E[Y|A=0,\boldsymbol{W}]) + \epsilon H(0,\boldsymbol{W})).$$
\\

\noindent Finally, we compute the ATE as the difference between expectations of the updated  $Q_{0}$ and $g_{0}$  predictions in the previous step (i.e., $\widehat{E}[Y|A=1,\boldsymbol{W}]) - \widehat{E}[Y|A=0,\boldsymbol{W}])$) and provide statistical inference using the functional Delta method.\citep{vanderLaan2011TargetedData, Luque-Fernandez2020DeltaTutorial}
\\

\noindent It is worth noting that Steps 3 and 4, which are improvements to AIPTW and IPTW-RA estimators, are the very concepts that make TMLE more robust against near positivity violations and force the estimator to respect the boundaries of the limits of the parameter space (ie., the probabilities stay between 0 and 1). For example, to estimate the ATE using the AIPTW estimator the researcher sets the estimation equation equal to zero. However, solving the estimating equation when there are near violations of the positivity assumption can cause the estimator to fall outside the boundaries of the parameter space (i.e., 0 and 1).
\\

\noindent The ATE estimate is 8.3\% (5.82 - 10.87) (Table 3), which is consistent with all the previous estimates using different estimators. Note that statistical inference is performed using the functional Delta method that is based on the Influence Function (IF).\citep{Vaart1998AsymptoticStatistics, Kennedy2016SemiparametricInference, vanderLaan2011TargetedData} \\

\subsection{Statistical inference for data-adaptive estimators: Functional Delta Method}
\noindent The IF is a fundamental object of semi-parametric theory that allows us to characterise a wide range of estimators and their efficiency.\citep{Vaart1998AsymptoticStatistics, Kennedy2016SemiparametricInference, vanderLaan2011TargetedData} The IF of a regular asymptotic and linear estimator 
$\hat{\psi}$ of $\psi(\theta)$, where $\theta$, is a random variable based on independent and identically distributed samples $ O_{i} $ which captures the first order asymptotic behaviour of $\hat{\psi}$, such that \\
$$ n^{\frac{1}{2}} {\hat{\psi}-\psi(\theta)} = n^{-\frac{1}{2}} \sum_{i=1}^{n} \textit{IF}(O_{i};\theta) + o_{p}(1).$$ \\
Where $o_{p}(1)$ represents the remainder term from the first order approximation that converges to zero in probability when the sample size converges to infinity. Mathematically we can identify the IF as being the second term of a first degree Taylor approximation.\citep{Luque-Fernandez2020DeltaTutorial} From the variance of the IF we derive the SE of the ATE from the TMLE estimator. Therefore, the functional Delta method based on the IF readily allows the application of the Central Limit Theorem and, therefore, to compute Wald-type confidence intervals.\citep{vanderLaan2011TargetedData} However, using the IF for statistical inference may require larger sample sizes to avoid finite-sample issues. Recent research and theoretical developments support the use of \text{doubly-robust cross fit} estimators (i.e., cross-validated) to improve statistical inference robustness based on the functional Delta method.\citep{Zivich2020MachineEstimators}
\\

\noindent In box 28, we outline how to compute the ATE using data-adaptive procedures implemented in the \textit{eltmle} user-written Stata command.\citep{Migariane/meltmle:Zenodo} \textit{eltmle} implements the TMLE framework for the ATE , the marginal RR and odds ratio for a binary, or continuous, outcome with a binary treatment. \textit{eltmle} includes the use of data-adaptive estimation of the propensity score $g_{0}$ and regression outcome $Q_{0}$ models via ensemble learning,\citep{VanDerLaan2007SuperLearner} which is implemented calling the \textit{SuperLearner} package v.2.0-21 from R.\citep{Polley2010SuperPrediction, VanDerLaan2007SuperLearner} The super-learner uses V-fold cross-validation (10-fold by default) to assess the performance of prediction regarding the potential outcomes and the propensity score as weighted averages of a set of machine learning algorithms. The \textit{SuperLearner} has default algorithms implemented in the base installation of the tmle-R package v.1.2.0-5.\citep{Gruber2011Tmle:Estimation} The default algorithms include the following: i) stepwise selection, ii) generalized linear modeling (GLM), iii) a GLM variant that includes second order polynomials and two-by-two interactions of the main terms included in the model. Additionally, \textit{eltmle} has an option to include Bayes Generalized Linear Models and Generalized Additive Models as additional Super-Learner algorithms. Future implementations will offer more advanced machine learning algorithms.\\

\noindent \textbf{Box 28: TMLE with data-adaptive estimation using the Stata's user writen \textit{eltmle}}
\begin{lstlisting}
    preserve
    eltmle $Y $A $W, tmle   // install via "ssc install eltmle" or "github install migariane/eltmle"  
    restore
\end{lstlisting}
\noindent The ATE is 8.3\% (5.82-10.87). We also obtain estimates of the causal risk ratio (CRR 1.28; 95\% CI 1.19-1.37) and the marginal odds ratio (MOR 1.45; 95\% CI 1.29-1.61). These, and previous, results are presented in Table 3.
\\

\section{Simulation}
The motivation of this section is to share with the reader the template needed to run a simple Monte Carlo simulated experiment to compare all the different methods provided in the tutorial. For simplicity we run one sample estimate. However, we provide the results and code in R of a Monte Carlo experiment with 1,000 samples based on the same template as the one presented here and available at \hyperlink{https://github.com/migariane/TutorialCausalInferenceEstimators}{https://github.com/migariane/TutorialCausalInferenceEstimators}. 
\\

\noindent In Box 29 we show the data generation process to create random variables including the confounders, the treatment and the outcome. Afterward, we estimate the simulated value for the ATE. Then, we compute the ATE using all the different estimators described above, and finally we compare their performance based on the relative bias with respect to the value of the simulated ATE. The variables generated in this simulation reflect the context of cancer survival epidemiology. This simulation includes a binary outcome (Y), potential outcomes i.e., Y(1) and Y(0) and a binary treatment (A). The confounders \textit{W} reflect the commonly analysed cancer patient characteristics: deprivation level (w1, five categories), age at diagnosis (w2), cancer stage (w3, four categories) and comorbidity (w4, four categories).\\

\noindent \textbf{Box 29: Data generation for the Monte Carlo experiment}\\
\begin{lstlisting}
	// Data generation
		clear
		set obs 1000
		set seed 777
		
	// Confounders
		gen w1 = round(runiform(1, 5)) //Quintiles of Socioeconomic Deprivation
		gen w2 = rbinomial(1, 0.45) //Binary: probability age >65 = 0.45
		gen w3 = round(runiform(0, 1) + 0.75*(w2) + 0.8*(w1)) //Stage 
		recode w3 (5/6=1) //Stage (TNM): categorical 4 levels
		gen w4 = round(runiform(0, 1) + 0.75*(w2) + 0.2*(w1)) //Comorbidites: categorical four levels
		
	// Binary treatmet
		gen A = (rbinomial(1,invlogit(-1 -  0.15*(w4) + 1.5*(w2) + 0.75*(w3) + 0.25*(w1) + 0.8*(w2)*(w4)))) 
	// Potential outcomes and simulated ATE (one run)
		gen Y1 = (invlogit(-3 + 1 + 0.25*(w4) + 0.75*(w3) + 0.8*(w2)*(w4) + 0.05*(w1))) // Potential outcome 1
		gen Y0 = (invlogit(-3 + 0 + 0.25*(w4) + 0.75*(w3) + 0.8*(w2)*(w4) + 0.05*(w1))) // Potential outcome 2
		gen psi = Y1-Y0  // Simulated ATE
		gen Y = A*(Y1) + (1 - A)*Y0 // Outcome
    
    // ATE from different estimators
    teffects ra (Y w1 w2 w3 w4) (A)     
    estimates store ra
    teffects ipw (Y) (A w1 w2 w3 w4)    
    estimates store ipw
    teffects ipwra (Y w1 w2 w3 w4) (A w1 w2 w3 w4)  
    estimates store ipwra
    teffects aipw (Y w1 w2 w3 w4) (A w1 w2 w3 w4)   
    estimates store aipw
    qui reg psi
    estimates store psi
    estout psi ra ipw ipwra aipw
    
    // ATE from Ensemble learning maximum likelihood estimation (ELTMLE)
    preserve
    eltmle Y A w1 w2 w3 w4, tmle    
    restore
			
	// Relative bias for the ATE 
    * Regression adjustment
      display abs(0.1787 -0.203419)/0.1787 
      0.1383268 // 13.8% bias
    * IPTW
      display abs(0.1787 - 0.2776)/0.1787 
      0.55344152 // 55.3% bias
    * IPTW-RA
      display abs(0.1787 - .2052088)/0.1787
      0.1483424 // 14.8% bias
    * AIPTW
      display abs(0.1787 - 0.2030)/0.1787
      0.13598209 // 13.6% bias
    * ELTMLE
      display abs(0.1787 - 0.1784)/0.1787
      0.00167879 // 0% bias
\end{lstlisting}

\noindent It is clear to see that, compared to the true ATE of 17.9\%, all of the methods (RA [13.8\%], IPTW [55.3\%], IPTW-RA [14.8\%], AIPTW [13.6\%]) produce a biased estimate, but ELTMLE produces an estimate that is unbiased relative to the true ATE. Note that the relative bias of IPTW is very large; IPTW relies on the positivity assumption, which in this simulation is violated because there was a low number of individuals with a higher comorbidity value. Without correcting for this imbalance in the data the methods that rely on this assumption will be vulnerable to bias.
\\

\section{Conclusion}

\noindent Overall, all the methods introduced here include the estimation of the G-formula (non-parametrically or parametrically), which is a generalization of standardization, and the inverse probability of treatment weighting (IPTW).\cite{Robins1986AEffect} However, there are other estimators based on matching strategies that we did not discuss in our tutorial.\citep{Rosenbaum1983TheEffects}\\

\noindent Table 3 shows all the ATE results for all the different estimators we introduced in the tutorial. Overall, all the methods showed a consistent result for the ATE. In a well-balanced data set, all of the methods used to address confounding consistently provide the same result (\ref{table:3}). The RHC data (demonstrated in this paper) is used to teach causal inference methods because of its extremely well balanced distribution of confounders across levels of the treatment (RHC). However, in most observational studies, data is not usually well balanced and there are potentially near violations of the positivity assumption that must always be checked. \\

\noindent We introduce the different estimators in regards to their chronological development; the methods were developed to answer the limitations of the previous approach. For example, parametric estimators were developed to answer the question of the curse of dimensionality of the non-parametric implementation of the G-formula. Then, issues related to extrapolation for the G-computation, and the instability of the estimations due to large weights for the IPTW estimators, encouraged the development of double-robust methods. AIPTW was a strong candidate to answer this issue by incorporating semi-parametric theory and methods to causal inference. However, researchers soon realised that it did not solve the estimation equation equal to zero due to the fact that it is not a substitution estimator or plug-in estimator (see glossary). Thus, to overcome this limitation of the AIPTW estimator, data-adaptive estimation using machine learning algorithms and ensemble learning to estimate the nuisance parameters from the regression and propensity score models, combined to a smart way to solve the estimation equation were developed later in time. Evidence shows that the double robust estimators (particularly TMLE) obtain less biased estimates of the true causal effect in comparison to naive estimators such as multivariate regression.\\

\noindent Evidence shows that comparing the underlying properties of each method based on Monte Carlo experiments, only TMLE provides the numerous properties of estimating the probability distribution that enable it to out-perform the others. The properties of the estimator are: loss-based, well-defined, unbiased, efficient and can be used as a substitution estimator. Maximum likelihood estimation (MLE) based methods (stratification, propensity score and parametric regression) and estimating equations (IPTW and AIPTW) do not have all of the properties of TMLE and will underperform in comparison. For more detailed comparisons between the different methods, we refer the reader to Chapter 6 of TMLE.\citep{vanderLaan2011TargetedData}. It is important to highlight that in contrast to the AIPTW estimator, TMLE respects the global constraints of the statistical model (i.e. $P_{0}(0<Y<1) = 1$) and solves the estimation equations being equal to zero.\\

\noindent However, even if TMLE is less prone to errors due to misspecification than alternative methods like inverse probability weighting, there is some question regarding the validity of the robustness of inference produced by TMLE in non-parametric settings. This is an area of ongoing work. Furthermore, TMLE is currently limited to analysis conducted in R, and the only current implementation of TMLE outside of R is the user written program \textit{eltmle} for Stata,\citep{Migariane/meltmle:Zenodo}. \textit{eltmle} is not completely native to Stata but rather calls the \textit{SuperLearner} R package to calculate the predictions of the exposure and outcome models. More work is required to implement TMLE in other statistical packages.\citep{Gruber2011Tmle:Estimation}\\

\noindent Causal inference is a growing field with many developments during the recent years. Modern causal inference methods allow machine learning to be used when strong assumptions for parametric models are not entirely valid or reasonable. Overall, due to the difficulty of correctly specifying parametric models in high dimensional data, we advocate for the use of doubly-robust estimators with ensemble learning. However, using these approaches may require larger sample sizes to avoid finite-sample issues. More importantly, the development of renewed statistical inference for estimators using data adaptive procedures given that there was not theoretical support for the use of the Bootstrap is now well established.\citep{Kennedy2016SemiparametricInference,Non-JSTOR} However, recent developments support the use of \text{doubly-robust cross fit} estimators to improve statistical inference robustness based on the functional Delta method.\citep{Zivich2020MachineEstimators} Tutorials introducing the use and derivation of the functional Delta method and Influence Curve for applied researchers are needed. The tutorial presented here may help applied researchers to gain a better understanding of computational implementation of different causal inference estimators from an applied perspective.\\

\begin{table}[ht!]
\centering
\caption{Results of the Average Treatment Effect (ATE) from the methods.}
\label{table:3}
\begin{threeparttable}
 \begin{tabular}{ l c c c c c } 
    & \multicolumn{2}{c}{\textbf{\textit{teffects}}} & & \multicolumn{2}{c}{\textbf{\textit{Bootstrap}}}  \\ 	\cline{2-3} \cline{5-6}
 \textbf{Method}    & \textbf{ATE} & \textbf{Naive 95\% CI} & & \textbf{ATE} & \textbf{BS 95\% CI} \\ 
 \hline
 &&&&& \\
 \multicolumn{1}{l}{\textbf{Univariable}} & & & & &  \\ \cline{1-1}
 Regression         & 7.4 & 4.84 -- 9.86    & & n/a   & n/a \\
 NPG - 1C           & n/a & n/a             & & 7.4   & 4.91 -- 10.01 \\
 NPG - FS           & 7.4 & 4.83 -- 9.91    & & n/a   & n/a   \\
 PG - 1C            & 7.4 & 4.83 -- 9.91    & & 7.4   & 4.73 -- 10.23 \\
 PG - FS            & 8.3 & 5.80 -- 10.85   & & 8.3   & 5.77 -- 10.83 \\
 &&&&& \\
 \multicolumn{1}{l}{\textbf{Multivariable}} & & & & &  \\ \cline{1-1}
 Regression         & 8.3 & 5.74 -- 10.72   & & n/a   & n/a  \\
 IPW - PS           & 8.3 & 5.78 -- 10.83   & & 8.3   & 5.71 -- 10.90 \\
 IPW - RA           & 8.3 & 5.79 -- 10.84   & & 8.3   & 5.76 -- 10.87 \\
 AIPW               & 8.3 & 5.79 -- 10.84   & & 8.3   & 5.91 -- 10.86 \\
 TMLE               & 8.3 & 5.82 -- 10.87   & & n/a   & n/a  \\
 ELTMLE             & 8.3 & 5.82 -- 10.87   & & n/a   & n/a  \\ [1ex] 
 \hline
 \end{tabular}
 \footnotesize{\textbf{NPG} = Non-Paramteric G-formula, \textbf{FS} = Fully saturated, \textbf{PG} = Parametric G-formula, \textbf{IPW} = Inverse Probability Weighting, \textbf{PS} = Propensity Score, \textbf{RA} = Regression Adjustment, \textbf{AIPW} = Augmented Inverse Probability Weighting, \textbf{TMLE} = Targeted Maximum Likelihood Estimation by hand, \textbf{ELTMLE} = Ensemble Learning Targeted Maximum Likelihood Estimation using Stata \textit{eltmle} package.}
 \end{threeparttable}
\end{table}

\section*{Funding}
\noindent MALF is supported by a Miguel Servet I Investigator Award (grant CP17/00206 EU-FEDER) from the National Institute of Health, Carlos III (ISCIII), Madrid, Spain. His funders had no role in the study design, data collection, data analysis, data interpretation, or writing of the report.

\section*{Authors contributions}
\noindent The article arises from the motivation to disseminate the principles of modern epidemiology amongst clinicians and applied researchers. MALF developed the concept, designed the first draft of the article and the computing code. All authors interpreted and reviewed the code and the data, drafted and revised the manuscript. All authors read and approved the final version of the manuscript. MALF is the guarantor of the article.

\section*{Acknowledgements}
\noindent We thank Fiona Ingleby and Cristina Renzi for their comments and testing the code included in the boxes of the article.

\section*{Glossary}
The glossary is adapted from the book \textit{"Targeted learning: causal inference for observational and experimental data"} and a recent publication introducing the TMLE framework.\citep{Coyle2020TargetingResearch, vanderLaan2011TargetedData}
\begin{itemize}
\item Data-generating process (DGP)\\
The true mechanism that generated the observed data, with the corresponding data-generating probability distribution which produces the observed samples that were collected.
\item Estimator Estimate \\
A function of the sample of observations (that is, a function of the random variables) that generates estimates.
The realized value of an estimator, or a function of the realized observations.
\item Counterfactual \\
A contrary-to-fact value said to arise from hypothetically imposing an intervention on a system represented by a structural causal model. For example, the potential outcome $Y(a)$ is a counterfactual that arises from a hypothetical intervention that sets the treatment A to level a .
\item Statistical model\\
A set (family) of probability distributions that could describe the data-generating process. Note that, outside simulation exercises, the true data-generating process is unknown.
\item Fully-saturated model \\
The data one would have observed in the ideal (impossible) experiment, and the set of possible probability distributions of the full-data random variable. In a causal model, the full data includes the counterfactual values of the outcome (i.e., potential outcomes) under all treatment/exposure conditions.
\item Model misspecification\\
A scenario in which the statistical model, which is postulated to contain the distribution describing the data-generating process, fails to actually contain the corresponding true data-generating distribution.
\item Parametric statistical model \\
A family of probability distributions indexed by a finite set of model parameters. For example, a linear model traditionally assumes the outcome is a linear function of covariates plus a normally distributed error term with constant variance. Its parameters are the coefficients on the covariates and the variance of the error term.
\item Non-parametric statistical model\\
 A family of probability distributions that cannot be indexed by a finite set of parameters. That is, the set of parameters indexing this family of distributions is infinite-dimensional. Most often, when making minimal assumptions, the data-generating process cannot be defined by a finite set of parameters, making the set of parameters infinite-dimensional. For example, if all we know about the data-generating process is that we have access to $n$ independent and identically distributed (i.i.d.) samples, then the statistical model for the data-generating process is a non-parametric statistical model.
\item Target estimand or target parameter\\
    A function of the true (unknown) data-generating process that one is interested in estimating, and represents the mathematical formulation of the motivating question of interest.
\item Maximum likelihood estimation \\
  The most common method for estimating parameters in a finite-dimensional model (i.e., parametric statistical model). As the name implies, such estimates are generated by finding a set of parameter estimates that maximize the likelihood function of the observed data.
\item Score equation \\
The gradient (i.e., multi-variable generalization of the derivative) of the log-likelihood function of the data with respect to the parameter(s). This equation provides information on the degree of change resulting from very small perturbations of the parameter values.
\item Regular estimator\\
A class of estimators that converge in distribution to some limit distribution even if one samples from a slightly perturbed data distribution. Such estimators, if also asymptotically linear, accommodate inference by way of their asymptotic convergence to a Normal distribution.
\item Plug-in (substitution) estimator\\
An estimator that generates an estimate of the true parameter value by “plugging in” estimates of relevant parts of the data-generating distribution into the parameter mapping. This method is commonly referred to as the plug-in principle. For example, “plugging in” targeted Super Learner fit of the conditional mean under A = 1 and A = 0 generates an estimate of the average treatment effect.
\end{itemize}

\bibliography{references.bib}

\end{document}